\documentclass[fleqn,usenatbib]{mnras}

\usepackage{newtxtext,newtxmath}

\usepackage[T1]{fontenc}

\DeclareRobustCommand{\VAN}[3]{#2}
\let\VANthebibliography\thebibliography
\def\thebibliography{\DeclareRobustCommand{\VAN}[3]{##3}\VANthebibliography}

\usepackage{graphicx}	
\usepackage{amsmath}	
\usepackage{pdflscape}
\usepackage{threeparttable}
\usepackage{multirow}
\usepackage{siunitx}
\usepackage{hyperref}
\usepackage{color}

\title[X-ray Analysis and Simulations of SMC X-1]{X-ray Analysis and Photon-transport Simulations of SMC X-1: A Warped-disc Origin of the Superorbital Modulation}

\author[S. Takashima et al.]{
S. Takashima, $^{1}$\thanks{E-mail: satoshi.takashima@riken.jp}
H. Odaka, $^{2,3}$
R. Tomaru, $^{2}$
A. Tanimoto, $^{4}$
A. Bamba, $^{5,6,7}$
T. Tamagawa, $^{1,8,9}$
\\
$^{1}$Cosmic Radiation Laboratory, RIKEN Nishina Center, 2-1 Hirosawa, Wako, Saitama 351-0198, Japan\\
$^{2}$Department of Earth and Space Science, Osaka University, 1-1 Machikaneyama, Toyonaka, Osaka 560-0043, Japan\\
$^{3}$Kavli IPMU, The University of Tokyo, Kashiwa 113-0033, Japan\\
$^{4}$Graduate School of Science and Engineering, Kagoshima University, 1-21-24, Korimoto, Kagoshima, Kagoshima 890-0065, Japan\\
$^{5}$Research Center for Early Universe, Faculty of Science, The University of Tokyo, 7-3-1 Hongo, Bunkyo-ku, Tokyo 113-0033, Japan\\
$^{6}$Department of Physics, Faculty of Science, The University of Tokyo, 7-3-1 Hongo, Bunkyo-ku, Tokyo 113-0033, Japan\\
$^{7}$Trans-Scale Quantum Science Institute, The University of Tokyo, Tokyo  113-0033, Japan\\
$^{8}$Department of Physics, Tokyo University of Science, 1-3 Kagurazaka, Shinjyuku-ku, Tokyo 162-8601, Japan\\
$^{9}$High Energy Astrophysics Laboratory, RIKEN Pioneering Research Institute, 2-1 Hirosawa, Wako, Saitama 351-0198, Japan
}

\date{Accepted 2026 January 19. Received 2026 January 18; in original form 2025 October 10}

\pubyear{\the\year{}}

\begin{document}
\label{firstpage}
\pagerange{\pageref{firstpage}--\pageref{lastpage}}
\maketitle

\begin{abstract}
The luminous accreting pulsar SMC X-1 is an appropriate target to explore the accretion dynamics. 
SMC X-1 shows unique quasi-periodic flux variability of 40--65\,days known as superorbital modulation. 
To constrain the accretion structure of SMC X-1 based on timing and spectral study, we have analysed X-ray data of SMC X-1 observed by Suzaku and NuSTAR at various epochs between 2011 and 2022. 
The spectral analysis shows that the hydrogen column density ($N_\mathrm{H}$) increases from $1.1 \times 10^{22}\,\mathrm{cm^{-2}}$ to $1.24 \times 10^{23}\,\mathrm{cm^{-2}}$ as the flux decreases with the superorbital modulation.
The neutral iron K$\alpha$ line at \SI{6.4}{keV} has a broad width of \SI{0.3}{keV}, and its equivalent width increases as toward superorbital low states. 
The line broadening is consistent with Keplerian motion at the inner disc rather than the stellar wind velocity of the donor star. 
These findings support that the superorbital modulation is a consequence of X-ray attenuation by the warped accretion disc. 
To test this interpretation, we have conducted photon transport simulations of a system consisting of a neutron star, a warped disc, and optically-thin disc atmosphere. 
Occultation of the central source by the disc successfully reproduces the observed variations in the equivalent width of neutral iron K$\alpha$ line, pulse profiles, and flux in hard X-rays. 
Notably, a disc precession angle of approximately \ang{30} can account for the observational features. 
For the radiation pattern of the photon source, the preferred beam width corresponds to a standard deviation of \ang{30}.
\end{abstract}




\section{Introduction}
\label{sec:introduction}
An accreting neutron star, accompanied by a normal donor star, is an excellent laboratory to study the physics of both strong magnetic and gravitational fields that cannot be reproduced in ground experiments.
Neutron stars in high-mass X-ray binaries (HMXBs) are particularly well-suited for such studies, since their magnetic fields can reach $\sim 10^{12}\,\mathrm{G}$.
However, the geometrical structure and physical conditions of the magnetized accretion flow remain poorly understood.
The key insights into these properties can be obtained from the time variability of X-ray emission from those systems.
Multiple observations are essential to probe the three-dimensional structure of the accretion system, which consists of accreted plasma on the magnetic poles, the infalling accretion flow toward the poles, and the large-scale accretion disc.
Three types of temporal variations are typically observed in HMXBs: those associated with the neutron star’s spin, orbital motion, and superorbital (SO) modulation.
The origin of the superorbital modulation, which is also observed in low-mass X-ray binaries \citep{Kotze2012} and ultraluminous X-ray sources (ULX) \citep{Bachetti2014}, remains mysterious.
Numerous hypotheses have been proposed, such as precession \citep{Whitehurst1991}, radiation-induced warping \citep{Petterson1975,Wijers1999}, and magnetic warping of an accretion disc \citep{Pfeiffer2004}.

The bright X-ray pulsar SMC X-1 \citep{Price1971} is crucial for constructing a unified picture of accretion geometries of HMXBs since this object shows all three types of the time variations with quasi-periodic superorbital modulation of 40--65 days \citep{Gruber1984,Trowbridge2007}.
This variable superorbital period makes SMC X-1 unique whereas many binaries including Her X-1 \citep{Giacconi1973} and LMC X-4 \citep{Lang1981} show stable superorbital periods.
SMC X-1 has an extremely high luminosity of $\sim 5\times 10^{38}\,\mathrm{erg\,s^{-1}}$, which moderately exceeds the Eddington limit.
This HMXB consists of a neutron star ($\mathrm{1.06\,M_{\odot}}$) \citep{VanderMeer2007} and a B0 Ib supergiant Sk 160 ($\mathrm{17.2\,M_{\odot}}$) \citep{Schreier1972}.
The binary shows pulsation of $0.71\,\mathrm{s}$ \citep{Lucke1976} and orbital modulation of 3.89 days \citep{Schreier1972}.

One of the clear pictures of the significant flux decrease in the superorbital modulation is obscuration of X-ray radiation on a line of sight between the neutron star and an observer by the warped and/or tilted accretion disc \citep{Wojdowski1998}.
We call this model a warped disc model in this paper.
In bright X-ray binaries, X-ray radiation from a central star can cause the accretion disc to tilt from the binary plane and to precess \citep{Katz1973}.
The warped disc model can explain not only both stability of tilt and precession of the disc, but also superorbital periods such as Her X-1, SS433, and LMC X-4 \citep{Wijers1999, Ogilvie2001}.
Note that the X-ray flux of SMC X-1 out of eclipse varies as high as 20 times while it varies few times in eclipse \citep{Clarkson2003}.
This is naturally explained if we suppose the following effect of blocking X-ray by the warped disc.
In eclipse, X-ray from the neutron star should be partly blocked by the companion star and the effect of blocking the X-ray by the warped accretion disc should get less effective.
The warped disc model might also hint at the mechanism of recently discovered ULX pulsars (ULXPs), as some of them demonstrate both superorbital modulation and super-Eddington luminosity \citep{Walton2016,Brightman2020}.

However, in the case of SMC X-1, the warped-disc model has not been confirmed.
\cite{Clarkson2003} analysed 1.3--100 keV data from  Burst And Transient Source Experiment (BATSE) aboard Compton Gamma-Ray Observatory and All-Sky Monitor (ASM) aboard Rossi X-ray Timing Explorer and found the coherent superorbital modulation in both the soft (1.3-12.1 keV) and hard(20-100 keV) bands.
They argued that the modulation reflects changes in the neutron star's intrinsic luminosity rather than obscuration by a warped disc since hard X-rays are far less influenced by absorption than soft X-rays. 
Subsequent monitoring, 10 Suzaku \citep{Mitsuda2007} observations in 2011–2012  (PI: J. Neilsen) and 19 NuSTAR \citep{Harrison2013} observations in 2012–2022, supports this view. 
\citet{Pradhan2020} reported consistently small equivalent widths of the neutral Fe K$\alpha$ line (10–270 eV) across these epochs, implying a direct view of the neutron star even during superorbital low states and indicating that at least part of the superorbital modulation arises from intrinsic luminosity variations.

To investigate the accretion geometry of SMC X-1, wide-band and long-term data obtained with pointed telescopes are required.
In this paper, we analyse all available Suzaku and NuSTAR data to examine the role of intrinsic luminosity variation of the neutron star.
The NuSTAR data include observations from 2016--2022, which were not covered in previous studies of SMC X-1 using the same instrument \citep{Pradhan2020,Brumback2020}.
We introduce a superorbital phase for each observation and examine how the spectrum and pulse shapes depend on this phase to clarify the origin of the superorbital modulation.
Additionally, we conduct a three-dimensional radiative transfer simulation that takes into account ionization process of atoms, which we can compare to the observed data to constrain the accretion geometry around SMC X-1.

The paper is organized as follows. 
Section~\ref{sec:observations} describes the observations and data reduction.
We present spectral analysis in Section~\ref{sec:spectral_analysis} and timing analysis in  Section~\ref{sec:timing_analysis}.
In Section~\ref{sec:monaco_analysis}, we conduct a radiative transfer simulation of an accreting neutron star with a warped accretion disc.
In Section~\ref{sec:discussion}, we discuss the accretion disc structure of SMC X-1 from the data analysis results, and give a favoured radiation distribution.
Section~\ref{sec:conclusions} gives our conclusions.

\section{Observations and Data Reduction}
\label{sec:observations}

\begin{figure*}
    \centering
    \includegraphics[height=8cm]{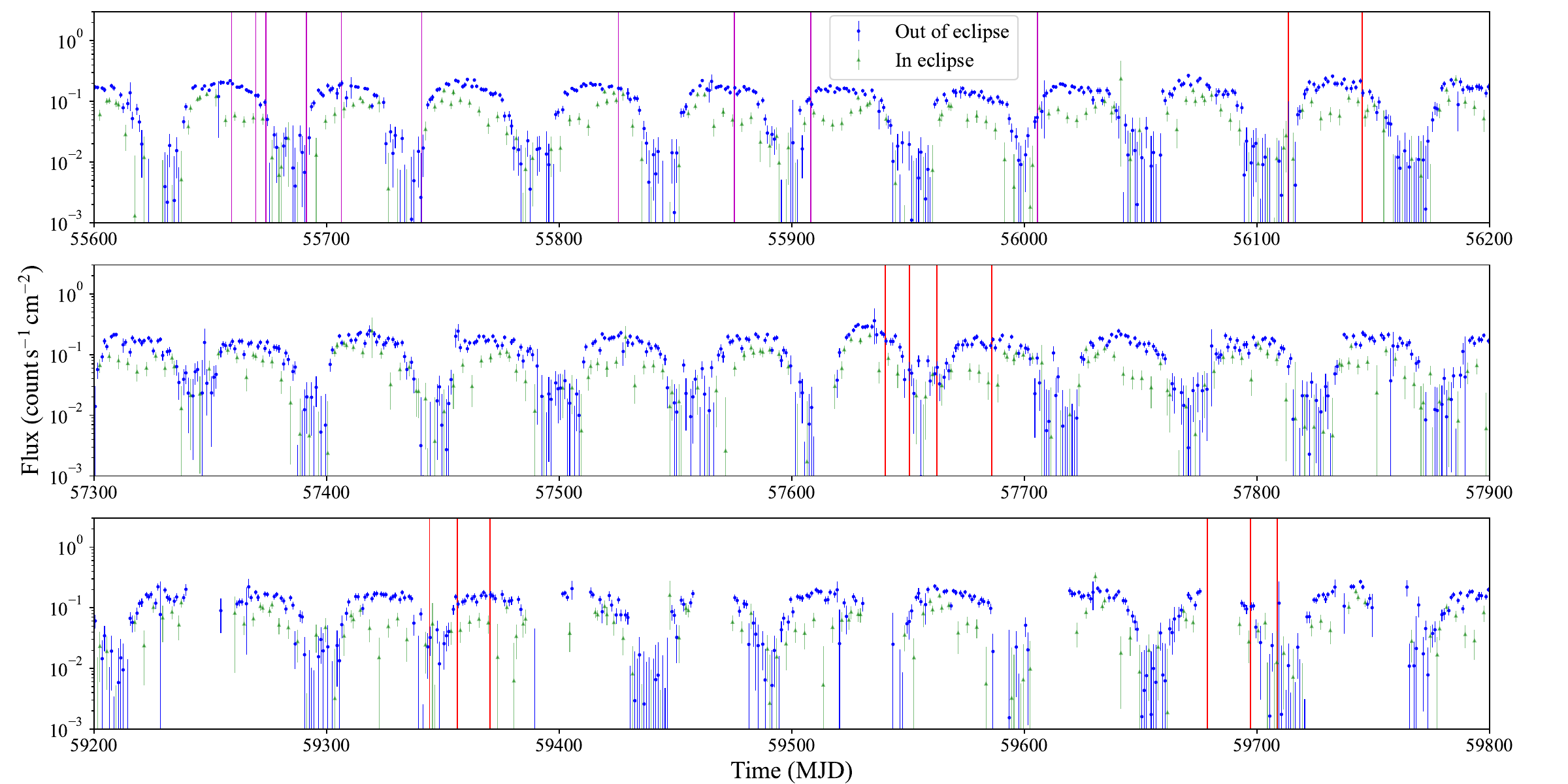}
    \caption{A light curve of SMC X-1 observed by MAXI (2--20\,keV). Blue data indicate observations out of eclipse ($0.15\leq \phi_\mathrm{orb} \leq 0.85$) while green ones, which appear in dips, correspond to the eclipse. Magenta (red) lines point to Suzaku (NuSTAR) observations.}
    \label{fig:lcurve}
\end{figure*}

In the following two subsections, we provide details of the Suzaku and NuSTAR observation data used in this analysis.
These observations were conducted during different phases of the quasi-periodic superorbital modulations, as illustrated in Fig.~\ref{fig:lcurve}.
This figure presents an X-ray light curve of SMC X-1 observed by the Monitor of All-sky X-ray Image (MAXI) \citep{Matsuoka2009} within the 2--20\,keV energy range. 
These MAXI data, binned in one-day intervals and processed as ver7L, exhibit clear superorbital modulations with a period of approximately 50 days. 
The epochs of Suzaku (magenta lines) and NuSTAR (red lines) observations of SMC X-1 are indicated on this light curve, demonstrating that the source was observed at various phases and cycles of the superorbital modulations.

\subsection{Suzaku}
Suzaku was equipped with four X-ray Imaging Spectrometers (XIS; \cite{Koyama2007}) and a Hard X-ray Detector (HXD; \cite{Takahashi2007}).
Suzaku performed 10 observations of SMC X-1 between April 2011 and March 2012, as summarized in Table \ref{tab:obsSuzaku}. 
The orbital phase $\phi_\mathrm{orb}$ is defined over the range $0\leq \phi_\mathrm{orb} < 1$, with $\phi_\mathrm{orb}=0$ corresponding to the mid-eclipse. 
For convenience, each observation ID is abbreviated using the last two or three digits followed by an "S" (e.g., S10, S20, ..., S100).

\begin{table}
    \centering
    \caption{Suzaku observations}
    \label{tab:obsSuzaku}
    \begin{tabular}{cccc}
        \hline
        ID          & Date$^{(\mathrm{a})}$ & Exposure & Orbital Phase$^{(\mathrm{b})}$\\  
        Full (Short)       &    & (ks)     & ($\phi_\mathrm{orb}$)\\
        \hline
        706030010 (S10)  & 2011-04-07 & 18.5 & 0.58--0.67\\
        706030020 (S20)  & 2011-04-18 & 17.3 & 0.24--0.36\\
        706030030 (S30)  & 2011-04-22 & 15.6 & 0.37--0.49\\
        706030040 (S40)  & 2011-05-10 & 17.9 & 0.84--0.94\\
        706030050 (S50)  & 2011-05-25 & 18.0 & 0.71--0.86\\
        706030060 (S60)  & 2011-06-28 & 18.7 & 0.58--0.70\\
        706030070 (S70)  & 2011-09-21 & 18.3 & 0.31--0.43\\
        706030080 (S80)  & 2011-11-10 & 19.9 & 0.13--0.27\\
        706030090 (S90)  & 2011-12-12 & 17.4 & 0.56--0.66\\
        706030100 (S100) & 2012-03-19 & 18.6 & 0.58--0.73\\ 
        \hline
    \multicolumn{4}{l}{$^{(\mathrm{a})}$ Start time of each observation, whose format is YYYY-MM-DD.}\\
    \multicolumn{4}{l}{\footnotesize$^{(\mathrm{b})}$ Orbital phases are defined such that $\phi_\mathrm{orb}=0$ at the mid-eclipse.}\\
    \end{tabular}
\end{table}

\begin{table}
    \centering
    \caption{NuSTAR observations}
    \label{tab:obsNuSTAR}
    \begin{tabular}{cccc}
        \hline
        ID            & Date  & Exposure & Orbital Phase\\  
        Full (Short)  &             & (ks)     & ($\phi_\mathrm{orb}$)\\
        \hline
        10002013001 (N1001) & 2012-07-05 & 36.6 & 0.34--0.60\\
        10002013003 (N1003) & 2012-08-06 & 18.6 & 0.52--0.60\\
        30202004002 (N3202) & 2016-09-08 & 22.5 & 0.61--0.73\\
        30202004004 (N3204) & 2016-09-19 & 21.1 & 0.28--0.40\\
        30202004006 (N3206) & 2016-10-01 & 20.4 & 0.29--0.41\\
        30202004008 (N3208) & 2016-10-24 & 20.8 & 0.41--0.53\\
        30702018002 (N3702) & 2021-05-10 & 26.8 & 0.48--0.60\\
        30702018004 (N3704) & 2021-05-21 & 29.3 & 0.56--0.68\\
        30702018006 (N3706) & 2021-06-04 & 34.8 & 0.14--0.26\\
        30702018008 (N3708) & 2022-04-09 & 32.8 & 0.40--0.54\\
        30702018010 (N3710) & 2022-04-27 & 33.5 & 0.17--0.31\\ 
        30702018012 (N3712) & 2022-05-09 & 41.2 & 0.13--0.27\\ 
        \hline
    \end{tabular}
\end{table}

In this work, we used the HEASARC software suite, \texttt{HEASoft} 6.31.1 including \texttt{Xspec} 12.13.0c for data preprocessing and spectral analysis \citep{Arnaud1996}.
We processed all Suzaku data using \texttt{aepipeline} to produce cleaned data that satisfied standard screening criteria.
When both $3\times 3$ and $5\times 5$ mode data were available in the same observation, we merged them using \texttt{XSELECT}.
We used calibration database version 20181010 for XIS and version 20110913 for HXD PIN.
Then, for spectral analysis of XIS, we selected X-ray events within a circle of $2'$ radius from the centre of the point source using \texttt{XSELECT} and \texttt{SAOImageDS9}.
We estimated background by using an off-source region.
In each spectrum, data points were binned such that each bin width of spectra was 60\,eV.
For the HXD spectral and timing analysis, non X-ray background (NXB) model was LCFITDT 2.0ver0804.
For pulse timing analysis, we used data only HXD PIN data as the XIS lacked sufficient time resolution for $\sim 0.7\,\mathrm{s}$ pulses.

\subsection{NuSTAR}
NuSTAR was launched in 2012 as the first space observatory equipped with a hard X-ray focusing system (3--79\,keV).
At the focal plane, two modules, FPMA and FPMB, each consisting of four CdZnTe pixel sensors, are installed.
Between 2012 and 2022, NuSTAR conducted 19 observations; we analysed 12 of these observations with exposure times longer than 1\,ks (see Table \ref{tab:obsNuSTAR}).
As in the case of Suzaku, we abbreviated the observation IDs, such as N1001 for 10002013001.
The NuSTAR data were processed with \texttt{nupipeline} and \texttt{nuproduct} to generate scientific-level event and spectral files.
For both spectral and timing analyses, source and background regions were extracted as two distinct circles of $90''$ radius, following the same procedure as for Suzaku.

\section{Spectral Analysis} \label{sec:spectral_analysis}
To evaluate the variability of X-ray spectra in the different structures of SMC X-1, we performed spectral analysis of 10 Suzaku and 12 NuSTAR observations.
The effective area of Suzaku is larger than that of NuSTAR in soft X-ray band up to $\sim 7\,\mathrm{keV}$, whereas NuSTAR provides better constraints on broadband spectral components.
Therefore, the spectral models applied to these data are not exactly the same, in the following parts.

\subsection{Suzaku}
\begin{figure}
    \centering
    \includegraphics[width=\columnwidth]{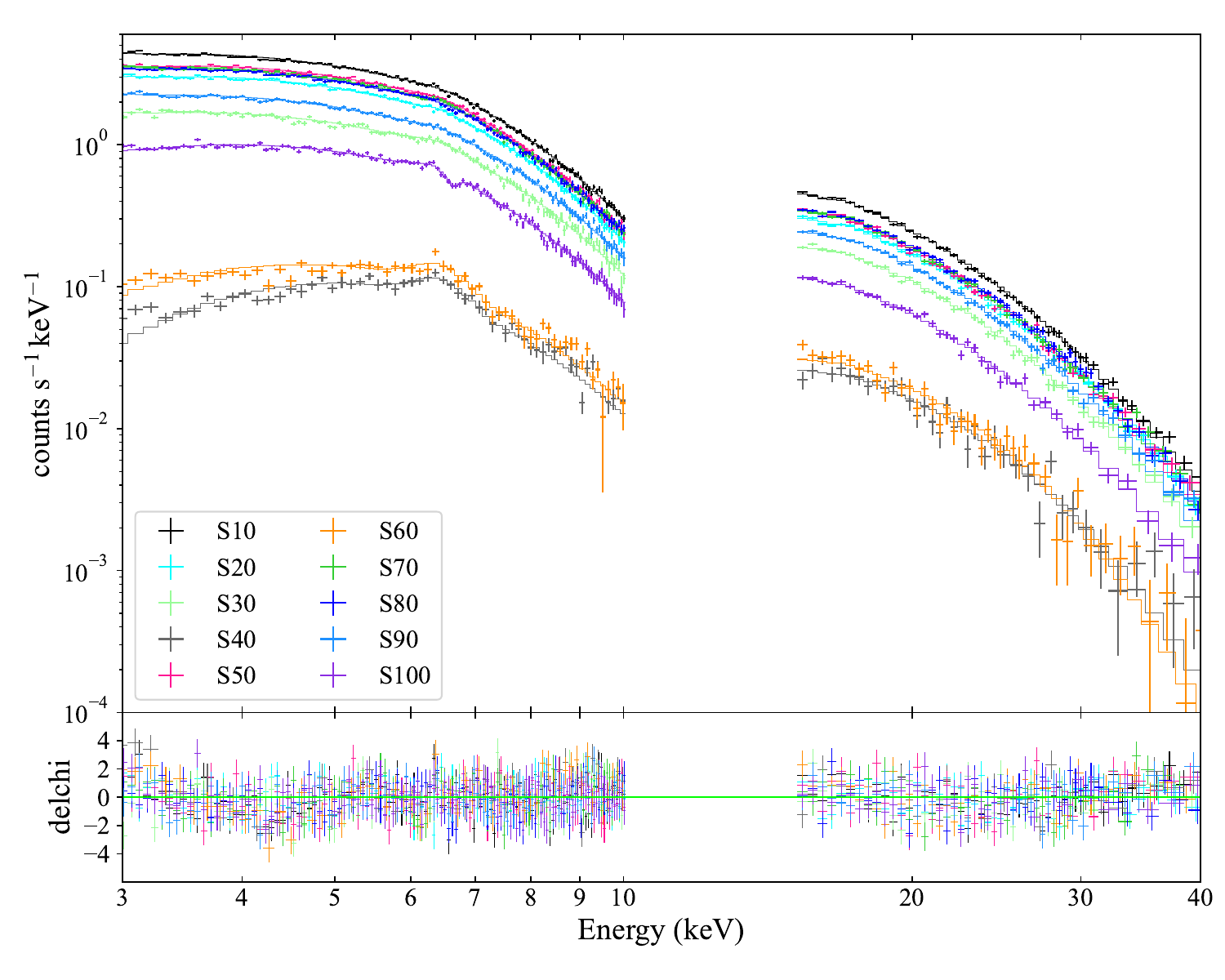}
    \caption{X-ray spectra of all the Suzaku SMC X-1 observations. Low energy region ($\leq\,10\,\mathrm{keV}$) and higher one ($\geq 15\,\mathrm{keV}$) correspond to data XIS0 and HXD PIN observed, respectively.}
    \label{fig:spectra_suzaku}
\end{figure}

We use an energy range of 3--10\,keV for XIS0 (a front-illuminated CCD) and a range of 3--6\,keV for XIS1 (a back-illuminated CCD) to avoid contamination from low-energy stellar wind emission.
The range of HXD PIN for the fitting was 15--40\,keV.
The spectra of the 10 observations by XIS0 (3--10\,keV) and HXD PIN (15--40\,keV) are plotted in Fig.~\ref{fig:spectra_suzaku}.
An emission line at 6.4\,keV, originating from neutral iron K$\alpha$ line, was detected in all the observations.
In two observations, S40 and S60, the spectrum shows suppression below \SI{5}{keV}, which is likely due to absorption.

We constructed a spectral model that contains the emission line and the absorption effect of circumambient materials step by step.
In the first step, we determined the continua of all 10 observations using data of XIS0, XIS1, and HXD PIN before evaluating the iron emission line.
We excluded the X-ray data in the energy range containing the neutral iron line when fitting the continuum.
We therefore decided not to use an energy range of 5.5--7.5\,keV for the two faintest observations, S40 and S60, and a range of 6.2--7.0\,keV for the brighter others.
We adopted the following continuum model in \texttt{Xspec}:
\begin{equation}
    \mathrm{constant}\langle 1 \rangle \ast \mathrm{tbabs}\langle 2 \rangle \ast f_{\mathrm{FDcut}}\langle 3 \rangle .
\end{equation}

The constant term was introduced to correct the small errors of the cross-calibration of the effective areas of XIS1 and HXD PIN compared with XIS0.
By definition, that term of XIS0 is fixed to unity.
For the constant value of HXD PIN, we followed the Suzaku ABC guide\footnote{https://heasarc.gsfc.nasa.gov/docs/suzaku/analysis/abc/abc.html} and fixed it to 1.16.
$\mathrm{tbabs}$ is an interstellar medium absorption model considering gas phase, molecules, and grain phase.
The abundance of elements followed \cite{Lodders2009}.
It is known that cutoff in a high energy region can reproduce the spectra of SMC X-1 \citep{Wojdowski1998}.
Thus, we adopted power-law model with a Fermi-Dirac cutoff \citep{Tanaka1986} which is widely used as a phenomenological radiation model of accreting neutron stars.
The modified power-law model can be written as follows:
\begin{equation}
    f_{\mathrm{FDcut}} = \frac{AE^{-1}}{1+e^{\left(E-E_\mathrm{c}\right)/E_\mathrm{f}}},
\end{equation}
where $A$ is a normalization factor, $E_\mathrm{c}$ and $E_\mathrm{f}$ are cutoff and folding energies, respectively.
While the original Fermi–Dirac cutoff model treats the photon index $\Gamma$ as a free parameter, many X-ray pulsars, including SMC X-1 and Her X-1, are known to exhibit continuum spectra with $\Gamma \sim 1$ \citep{Hu2025, Vasco2013}.
In practice, fixing $\Gamma = 1$ is often adopted in spectral analyses to avoid strong degeneracy among parameters.
Therefore, in this work we fixed the photon index at $\Gamma = 1$.
Any fitting result returned an acceptable reduced chi-squared of 1.10-1.73.
$N_\mathrm{H}$, $E_\mathrm{c}$ and $E_\mathrm{f}$ show moderate correlation (see Appendix. \ref{sec:corner_plot}).
Then, we calculated a flux in the range of 2--10\,keV.
The fluxes of S40 and S60, which correspond to observations of the low superorbital phases, are an order of magnitude smaller than those of other observations.
A large $N_\mathrm{H}$ was measured in these two superorbital low states, S40 ($(12.4 \pm 0.8) \times 10^{22}\,\mathrm{cm^{-2}}$) and S60 ($(7.4 \pm 0.6) \times 10^{22}\,\mathrm{cm^{-2}}$), the intermediate phase of S100 shows $(3.0\pm 0.3) \times 10^{22}\,\mathrm{cm^{-2}}$, other brighter observations showed small values of $\sim  10^{22}\,\mathrm{cm^{-2}}$.

\begin{landscape}
  \begin{table}
    \centering
    \caption{Parameters of spectral fitting $^{\ast}$}
    \begin{tabular}{ccccccccccc}
      \hline
        ID & S10 & S20 & S30 & S40 & S50 & S60 & S70 & S80 & S90 & S100$^{(\mathrm{f})}$ \\
      \hline
      SO phase$^{(\mathrm{a})}$
        & 0.54 & 0.72 & 0.79 & 0.15 & 0.48 & 0.16 & 0.66 & 0.60 & 0.23 & 0.19 \\
      $N_\mathrm{H}$\ ($10^{22}\,\mathrm{cm^{-2}}$) &
        $1.1 \pm 0.1$ & $1.4 \pm 0.2$ & $1.3 \pm 0.2$ & $12.4 \pm 0.8$ & $1.6^{+0.2}_{-0.1}$ & $7.4 \pm 0.6$ & $1.2 \pm 0.1$ & $1.2 \pm 0.1$ & $0.9 \pm 0.2$ & $3.0 \pm 0.3$\\
      $E_\mathrm{c}$\ (keV)  &
        $18 \pm 1$ & $17 \pm 1$ & $20^{+1}_{-2}$ & $25 \pm 1$ & $16 \pm 1$ & $24.9 \pm 0.8$ & $16 \pm 1$ & $18 \pm 1$ & $21 \pm 1$ & $17 \pm 2$\\
      $E_\mathrm{f}$ (keV)  &
        $8.0 \pm 0.3$ & $8.6 \pm 0.4$ & $7.9 \pm 0.5$ & $6^{+2}_{-1}$ & $8.3 \pm 0.3$ & $4.5^{+0.9}_{-0.7}$ & $8.4 \pm 0.3$ & $8.1 \pm 0.3$ & $7.3 \pm 0.4$ & $7.9 \pm 0.6$\\
      Flux $^{(\mathrm{b})}$ &
        $9.66^{+0.04}_{-0.07}$ & $6.66^{+0.03}_{-0.06}$ & $3.86^{+0.02}_{-0.05}$ & $0.283^{+0.005}_{-0.006}$ & $7.92^{+0.03}_{-0.04}$ & $0.376^{+0.006}_{-0.008}$ & $7.87^{+0.03}_{-0.08}$ & $7.81^{+0.03}_{-0.07}$ & $4.97^{+0.02}_{-0.03}$ & $2.34^{+0.01}_{-0.03}$\\
      constant (XIS1)       &
        $1.039 \pm 0.006$ & $1.002 \pm 0.008$ & $1.05 \pm 0.01$ & $ 1.0196^{(\mathrm{c})} $ & $1.020 \pm 0.006$ & $ 1.0196^{(\mathrm{c})} $ & $1.022 \pm 0.006$ & $1.041 \pm 0.006$ & $1.047 \pm 0.008$ & $1.02 \pm 0.01$\\
      $\chi_\nu^2$ $^{(\mathrm{d})}$ (d.o.f)        &
        1.42 (203) & 1.02 (201) & 1.22 (199) & 1.77 (115) & 1.54 (200) & 1.77 (119) & 1.24 (202) & 1.18 (204) & 1.02 (200) & 1.15 (192)\\
      $\sigma_{\mathrm{Fe}}$ (keV) &
        $0.40^{+0.2}_{-0.14}$ & $0.219^{+0.15}_{-0.084}$ & $0.3^{+0.14}_{-0.1}$ & $0.33^{+0.26}_{-0.13}$ & $0.41^{+0.14}_{-0.11}$ & $0.34 \pm 0.13$ & $0.32^{+0.16}_{-0.14}$ & $0.4^{+0.12}_{-0.1}$ & $0.31^{+0.14}_{-0.12}$ & $0.33^{+0.11}_{-0.09}$\\
      $\mathrm{EW_{Fe}}$ (eV) &
        $48^{+13}_{-12}$ & $39^{+10}_{-12}$ & $88 \pm 22$ & $276^{+63}_{-54}$ & $68^{+12}_{-14}$ & $316^{+44}_{-62}$ & $65^{+12}_{-13}$ & $56^{+15}_{-14}$ & $54^{+14}_{-18}$ & $120^{+31}_{-33}$\\
      $\chi_\nu^2$ $^{(\mathrm{e})}$ (d.o.f)       &
        1.25 (48) & 0.73 (48) & 0.979 (48) & 0.754 (23) & 1.2 (48) & 1.36 (23) & 1.1 (48) & 1.02 (48) & 0.691 (48) & 0.997 (45)\\
      \hline
      \multicolumn{11}{l}{$^{\ast}$ Errors of this work denote 90\% confidence levels.}\\
      \multicolumn{11}{l}{$^{(\mathrm{a})}$  Derived in Section \ref{sec:analysis_superorbital}.}\\
      \multicolumn{11}{l}{$^{(\mathrm{b})}$  in the unit of $10^{-10}{\,\mathrm{erg\,cm^{-2} s^{-1}}}$}\\
      \multicolumn{11}{l}{$^{(\mathrm{c})}$  Fixed parameters.}\\
      \multicolumn{11}{l}{$^{(\mathrm{d})}$  Fit without the iron K$\alpha$ line to determine best fit parameters of the continuum. }\\
      \multicolumn{11}{l}{$^{(\mathrm{e})}$ Fit with the iron K$\alpha$ line to determine the best fit parameters.}\\
      \multicolumn{11}{l}{$^{(\mathrm{f})}$ Multiplicative absorption model \textit{gabs} is also adopted in this model (see the main texts for more detail).}\\
      \multicolumn{11}{l}{\ \ \ \ \ \ \,Fit parameters of that \textit{gabs} is $6.62^{+0.03}_{-0.04}$\,keV (line energy), $0.08^{+0.04}_{-0.05}$\,keV (line width), $0.06 \pm 0.02$\,keV (line depth).}
    \end{tabular}\label{tab:fit_param_suzaku}
  \end{table}
\end{landscape}

In the second step, we added a neutral iron K$\alpha$ line with a Gaussian function and put those results.
The line energy was fixed at \SI{6.4}{keV}, except for the S100, where a neighboring absorption feature was present.
An absorption model, \textit{gabs}, was incorporated into only S100 data followed by the report of helium-like iron absorption line \citep{Kubota2018}.
The fitting was performed for all observations and yielded acceptable chi-squared values.
Fit parameters are listed in Table \ref{tab:fit_param_suzaku}.
Equivalent widths (abbreviated to $\mathrm{EW_{Fe}}$ in the table) of the neutral iron K$\alpha$ line are listed as well.

\subsection{NuSTAR}
\begin{figure}
    \centering
    \includegraphics[width=\columnwidth]{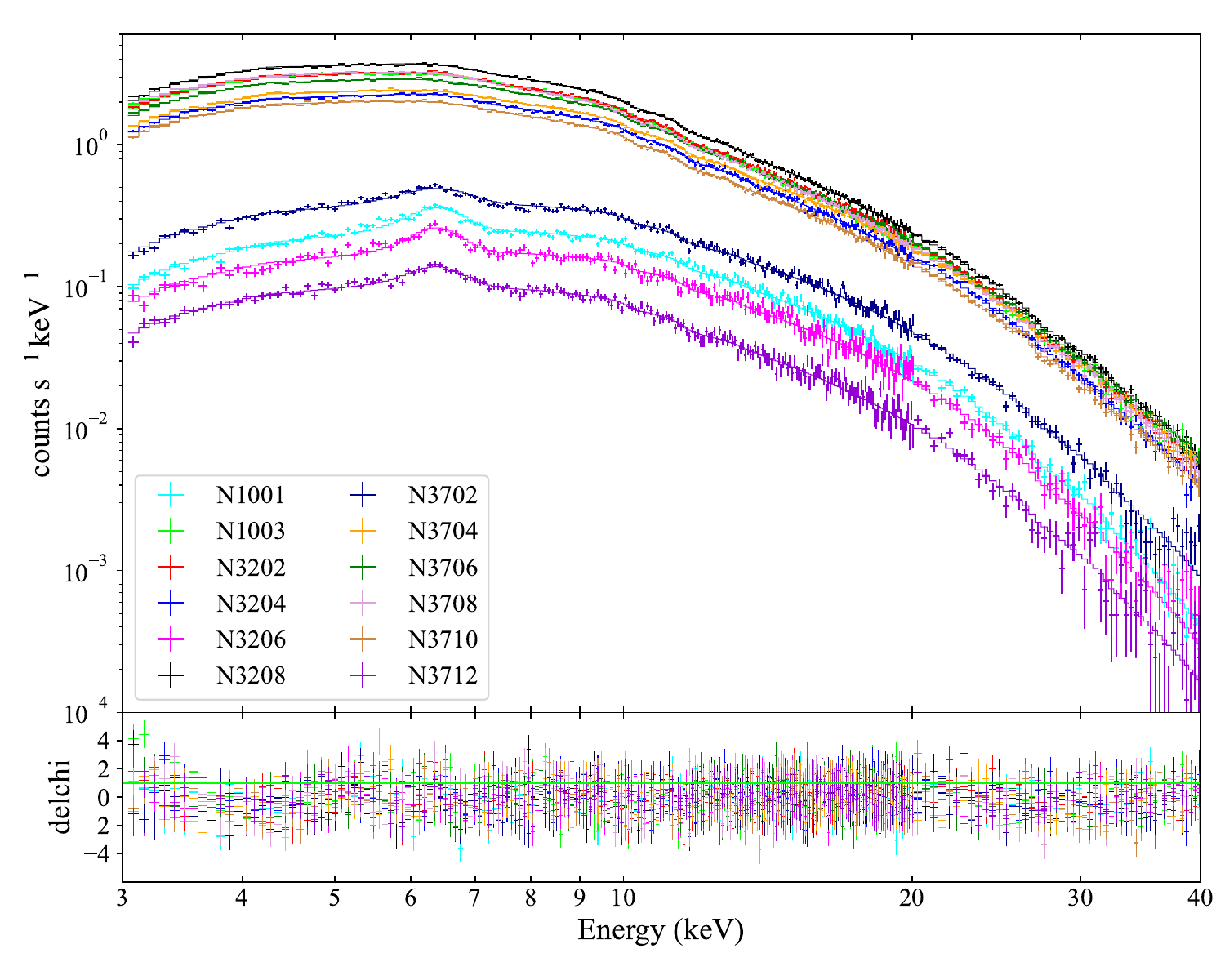}
    \caption{X-ray spectra from all NuSTAR observations and the best-fit models are shown in the top panel. The bottom panel displays the residuals, i.e., the difference between the data and the model normalized by the statistical uncertainty.}
    \label{fig:spectra_nustar}
\end{figure}

The 12 NuSTAR spectra were fitted with a model consisting of a continuum component plus a Gaussian line for the neutral iron K$\alpha$ emission (see Fig.~\ref{fig:spectra_nustar}). 
For the continuum of the NuSTAR data, the NPEX model \citep{Makishima1999,Odaka2013} was preferred over a power-law model with a Fermi-Dirac cutoff due to the lower fit residuals achieved.
The NPEX model is defined as $(E^{-\alpha} + BE^{-\beta})\exp(-E/E_\mathrm{c})$, where the parameter $\beta$ was fixed at -2 in this analysis.

In summary, the spectral fitting model applied to the NuSTAR data is represented as follows:
\begin{equation}
    \mathrm{constant}\langle 1 \rangle \times \mathrm{tbabs}\langle 2 \rangle \times (\mathrm{cflux}\langle 3 \rangle \times \mathrm{NPEX}\langle 4 \rangle+ \mathrm{gauss}\langle 5 \rangle).
\end{equation}
To determine the flux in the 3--40\,keV energy range, a \texttt{cflux} component was included as a multiplicative factor to the NPEX model. 
Due to the limited effective area of NuSTAR at lower energies, the absorption column density was fixed at $4.5\times 10^{21}\,\mathrm{cm^{-2}}$, a value derived from the HI4PI survey \citep{HI4PI2016} for the line of sight towards SMC X-1. 
Furthermore, the central energy of the neutral iron K$\alpha$ emission line was fixed at 6.4\,keV.

Fig.~\ref{fig:spectra_nustar} presents the NuSTAR observational data along with the best-fitting model and the corresponding residuals in the bottom panel. 
Four distinct low-flux state observations (N1001, N3206, N3702, N3712) are evident and consistent with the trends in the MAXI light curve (Fig.~\ref{fig:lcurve}).
Fit parameters are presented in Table \ref{tab:fit_param_nustar}.

\begin{landscape}
    \begin{table}
    \centering
    \caption{Fit parameters of the NuSTAR analysis}
    \begin{tabular}{@{}ccccccc@{}} \hline
    ID & N1001 & N1003 & N3202 & N3204 & N3206 & N3208\\ \hline \hline
    SO phase & 0.12 & 0.64 & 0.66 & 0.86 & 0.06 & 0.46 \\
    Flux (FPMA)$^{(\mathrm{a})}$ & $1.10 \pm 0.01$ & $8.94 \pm 0.06$ & $8.65 \pm 0.06$ & $6.81 \pm 0.05$ & $0.81 \pm 0.01$ & $9.91 \pm 0.05$\\
    $\mathrm{EW}$ (FPMA)$^{(\mathrm{b})}$ & $450 \pm 30$ & $80 \pm 10$ & $100 \pm 10$ & $80 \pm 10$ & $450 \pm 50$ & $76 \pm 9$\\
    Flux (FPMB)$^{(\mathrm{a})}$ & $1.09 \pm 0.01$ & $9.18 \pm 0.06$ & $8.45 \pm 0.06$ & $6.66 \pm 0.05$ & $0.79 \pm 0.02$ & $9.74 \pm 0.05$\\
    $\mathrm{EW}$ (FPMB)$^{(\mathrm{b})}$ & $460 \pm 50$ & $66.9^{+23.3}_{-0.8}$ & $90 \pm 10$ & $85 \pm 13$ & $440 \pm 40$ & $87 \pm 20$\\
    $\log{F_{3-40\,\mathrm{keV}}}$ & $-9.694 \pm 0.003$ & $-8.715 \pm 0.001$ & $-8.732 \pm 0.001$ & $-8.856 \pm 0.002$ & $-9.834 \pm 0.005$ & $-8.673 \pm 0.001$\\
    $\alpha$ & $0.05 \pm 0.07$ & $0.417 \pm 0.008$ & $0.406 \pm 0.007$ & $0.39 \pm 0.01$ & $0.3 \pm 0.1$ & $0.393 \pm 0.007$\\
    $B$ & $0.0146 \pm 0.0007$ & $0.00088 \pm 0.00008$ & $0.00107 \pm 0.00007$ & $0.00160 \pm 0.00009$ & $0.012 \pm 0.001$ & $0.00111 \pm 0.00007$\\
    $E_\mathrm{c}$ (keV) & $4.45 \pm 0.07$ & $5.76 \pm 0.09$ & $5.60 \pm 0.07$ & $5.40 \pm 0.06$ & $4.40 \pm 0.08$ & $5.58 \pm 0.07$\\
    $\sigma_\mathrm{G}$ (keV) & $0.36 \pm 0.03$ & $0.6 \pm 0.1$ & $0.55 \pm 0.09$ & $0.43 \pm 0.08$ & $0.32 \pm 0.04$ & $0.5 \pm 0.1$\\
    $A_\mathrm{G}$ & $0.00042 \pm 0.00003$ & $0.0009 \pm 0.0002$ & $0.0011 \pm 0.0001$ & $0.0006 \pm 0.0001$ & $0.00029 \pm 0.00003$ & $0.0010 \pm 0.0002$\\
    constant (FPMB) & $0.994 \pm 0.008$ & $1.027 \pm 0.003$ & $0.977 \pm 0.003$ & $0.978 \pm 0.004$ & $0.98 \pm 0.01$ & $0.984 \pm 0.003$\\ \hline
     & & & & & & \\ \hline
    ID & N3702 & N3704 & N3706 & N3708 & N3710 & N3712 \\ \hline \hline
    SO phase  & 0.96 & 0.27 & 0.51 & 0.46 & 0.76 & 0.91 \\
    Flux (FPMA)$^{(\mathrm{a})}$ & $2.04 \pm 0.02$ & $8.24 \pm 0.04$ & $9.14 \pm 0.04$ & $8.64 \pm 0.04$ & $6.36 \pm 0.03$ & $0.420 \pm 0.007$ \\ 
    $\mathrm{EW}$ (FPMA)$^{(\mathrm{b})}$ & $300^{+30}_{-20}$ & $90 \pm 10$ & $110 \pm 10$ & $150 \pm 20$ & $70 \pm 10$ & $360 \pm 50$ \\
    Flux (FPMB)$^{(\mathrm{a})}$ & $2.06 \pm 0.02$ & $8.31 \pm 0.05$ & $9.20 \pm 0.04$ & $8.63 \pm 0.05$ & $6.39 \pm 0.03$ & $0.423 \pm 0.007$ \\
    $\mathrm{EW}$ (FPMB)$^{(\mathrm{b})}$ & $290^{+40}_{-10}$ & $79 \pm 7$ & $110 \pm 20$ & $132 \pm 2$ & $70 \pm 10$ & $311 \pm 5$ \\
    $\log{F_{3-40\,\mathrm{keV}}}$ & $-9.451 \pm 0.003$ & $-8.798 \pm 0.001$ & $-8.739 \pm 0.001$ & $-8.728 \pm 0.001$ & $-8.892 \pm 0.001$ & $-10.101 \pm 0.005$ \\
    $\alpha$ & $0.17 \pm 0.06$ & $0.401 \pm 0.009$ & $0.413 \pm 0.008$ & $0.455 \pm 0.007$ & $0.414 \pm 0.009$ & $0.11 \pm 0.08$ \\
    $B$ & $0.0096 \pm 0.0004$ & $0.00159 \pm 0.00007$ & $0.00149 \pm 0.00006$ & $0.00109 \pm 0.00006$ & $0.00144 \pm 0.00008$ & $0.0091 \pm 0.0008$\\
    $E_\mathrm{c}$ (keV) & $4.80^{+0.07}_{-0.06}$ & $5.60 \pm 0.06$ & $5.53 \pm 0.05$ & $5.50 \pm 0.06$ & $5.52 \pm 0.06$ & $4.6 \pm 0.1$\\
    $\sigma_\mathrm{G}$ (keV) & $0.44\pm 0.05$ & $0.7 \pm 0.1$ & $0.76 \pm 0.09$ & $0.82 \pm 0.09$ & $0.6 \pm 0.2$ & $0.35 \pm 0.05$\\
    $A_\mathrm{G}$ & $0.00046 \pm 0.00004$ & $0.0007 \pm 0.0001$ & $0.0011 \pm 0.0001$ & $0.0017 \pm 0.0002$ & $0.0005 \pm 0.0001$ & $0.00014 \pm 0.00001$\\
    constant (FPMB) & $1.009 \pm 0.007$ & $1.007 \pm 0.003$ & $1.006 \pm 0.003$ & $0.998 \pm 0.003$ & $1.004 \pm 0.003$ & $1.01 \pm 0.01$\\
    \hline
    \multicolumn{7}{l}{$^{\ast}$ Errors of this work denote 90\% confidence levels.}\\
    \multicolumn{7}{l}{$^{(\mathrm{a})}$ Unit is $\mathrm{erg/cm^2/s}$}\\
    \multicolumn{7}{l}{$^{(\mathrm{b})}$ in the unit of eV}
    \end{tabular}\label{tab:fit_param_nustar}
    \end{table}
\end{landscape}

\section{Temporal Analysis} 
\label{sec:timing_analysis}
Studying the time variation of the spectral parameters with respect to the superorbital modulation is directly related to the accretion structure.
In Section \ref{sec:analysis_superorbital}, we define the superorbital phase to determine superorbital variability of the spectral parameters.
In Section \ref{sec:timing_analysis_pulse}, we analyse pulsation due to the neutron star's spin using data from both Suzaku HXD-PIN and NuSTAR.

\subsection{Superorbital Phase Analysis} \label{sec:analysis_superorbital}
Here, we examine the time variability of the spectral parameters described in Section \ref{sec:spectral_analysis} in relation to the superorbital modulation.
To evaluate this variability, it is convenient to define phase.
Unlike the orbital modulation, however, the superorbital modulation of SMC X-1 is not strictly periodic; as mentioned in Section \ref{sec:introduction}, its period can vary even between two successive cycles.
Therefore, it is helpful to represent such a modulation with a single cosine function as follows:
\begin{equation}\label{eq:variable_cosine}
    s(t)=A(t)\cos{\theta(t)},
\end{equation}
where $A(t)$ and $\theta(t)$ are the instantaneous amplitude and phase, respectively. 
Note that the phase is not simply proportional to time.
By applying the Hilbert transform to Equation~\ref{eq:variable_cosine}, one obtains
$\hat{s}(t)=A(t)\sin{\theta(t)}$, which is delayed by $-\pi/2$ with respect to $s(t)$, as expressed by
\begin{equation}\label{eq:hilbert_transform}
\hat{s}(t) = \frac{1}{\pi}\int_{-\infty}^{\infty}\frac{s(\tau)}{t-\tau}\mathrm{d}\tau.
\end{equation}
Once the superorbital component is represented by the above form, the phase is defined as follows:
\begin{equation}\label{eq:def_so_phase}
    \phi_\mathrm{SO}=\frac{\theta(t)}{2\pi}  = \frac{1}{2\pi}\arctan{\frac{\hat{s}(t)}{s(t)}}.
\end{equation}

Since the light curve of SMC X-1 contains not only an offset but also other frequency components such as the orbital modulation, it is required to extract components originating solely from the superorbital modulations.
Hilbert-Huang transform (HHT), proposed by \cite{Huang1998}, is a tool for interpreting time-series data as a stacked waveform of functions with time-dependent amplitudes and phases.
\cite{Hu2011} applied HHT to determine the superorbital phase of SMC X-1 using RXTE data.
To extract information on the instantaneous phase and amplitude of temporal data, the signal must be described by an intrinsic mode function (IMF).
We decompose the light curve into IMFs by using empirical mode decomposition (EMD) employed in HHT.

\begin{figure}
    \centering
    \includegraphics[width=9cm]{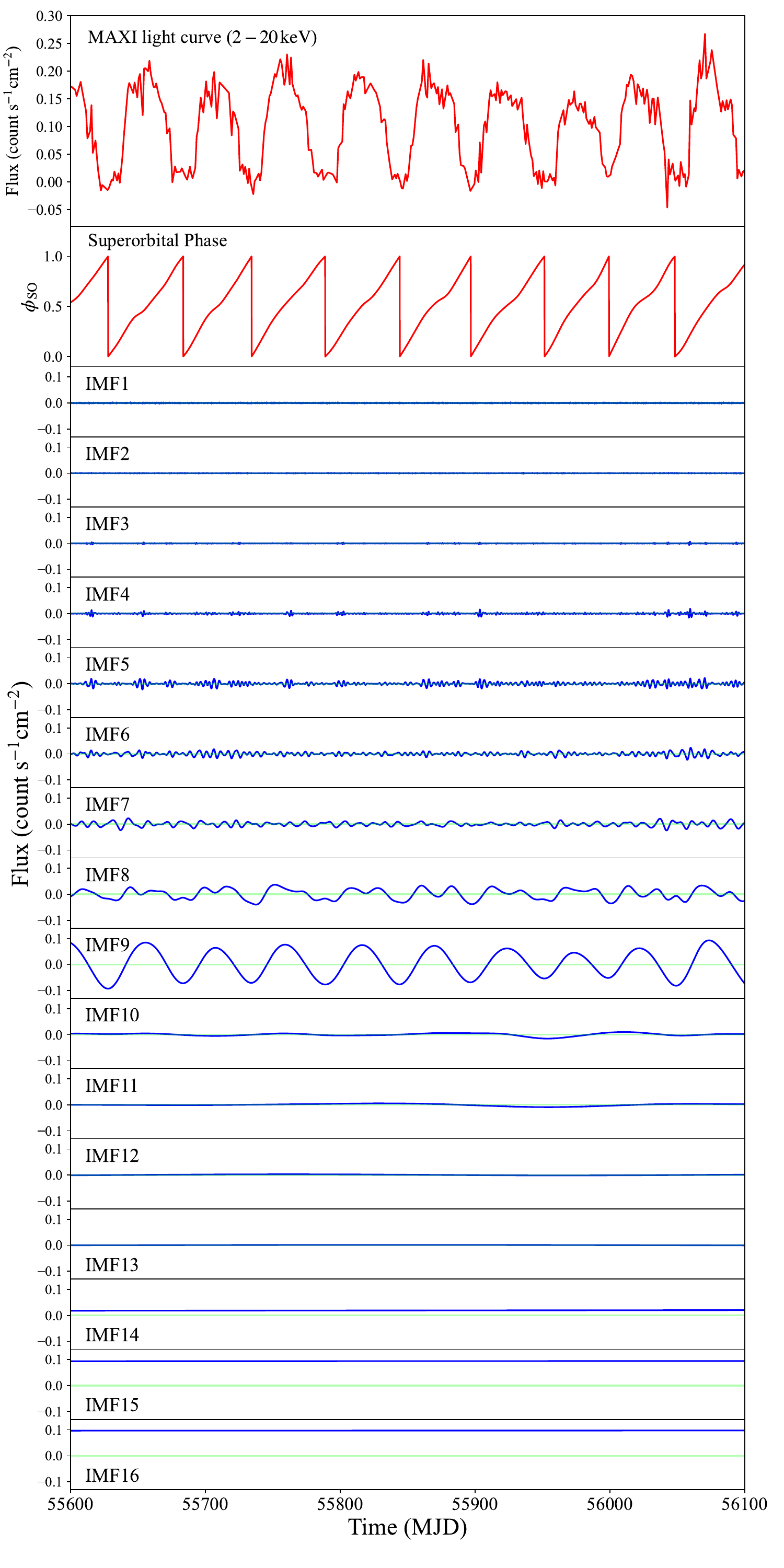}
    \caption{MAXI light curve (top), superorbital phases (second), and all 16 IMFs (the other panels) between MJD 55600 and 56100.}
    \label{fig:hht_analysis}
\end{figure}

We performed HHT to MAXI data (2--20\,keV) as the following three steps.
First, we interpolated the MAXI data to obtain evenly-sampled timing data.
To avoid being affected by the orbital modulation, data points corresponding to the orbital phase of $-0.15 < \phi_\mathrm{orb} < 0.15$ were eliminated (see Section \ref{sec:observations}).
The second step is decomposition into IMFs using ensemble EMD (EEMD), which repeats adding different white noises into input data many times, performing EMD, and then averaging the multiple outputs.
Figure~\ref{fig:hht_analysis} shows the MAXI light curve excluding eclipse data (top panel), the instantaneous superorbital phase (second panel), and all 16 IMFs between MJD 55600 and 56100 (remaining panels).
Decomposed IMFs sometimes show close frequency modulation (mode mixing), though EEMD is known to mitigate mode mixing.
Therefore, a pair of decomposed IMFs was merged into one function when the correlation function of the two IMFs of that pair was greater than 10.0.
Among the 16 IMFs, IMF 8 and 9 show such a significant correlation.
Because they show time variations with periods of 45--60 days, we summed them and regarded the combined signal as the superorbital component.
Then, we obtained the superorbital modulation curve as an IMF, which enables us to define the superorbital phase $\phi_\mathrm{SO}$ by Equation~\ref{eq:def_so_phase}.

\begin{figure}
    \centering
    \includegraphics[width=\columnwidth]{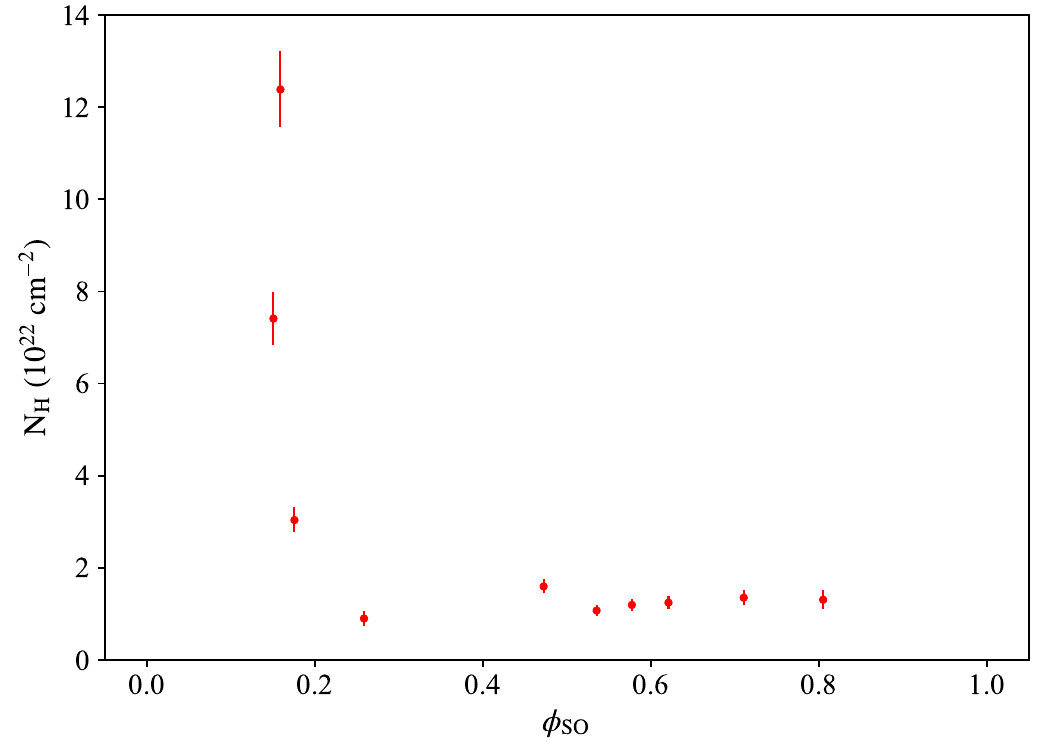}
    \caption{Dependence of the hydrogen column density derived from the Suzaku spectral analysis on the superorbital phase ($\phi_\mathrm{SO}$).}
    \label{fig:so_vs_nh}
\end{figure}

\begin{figure}
    \centering
    \includegraphics[width=\columnwidth]{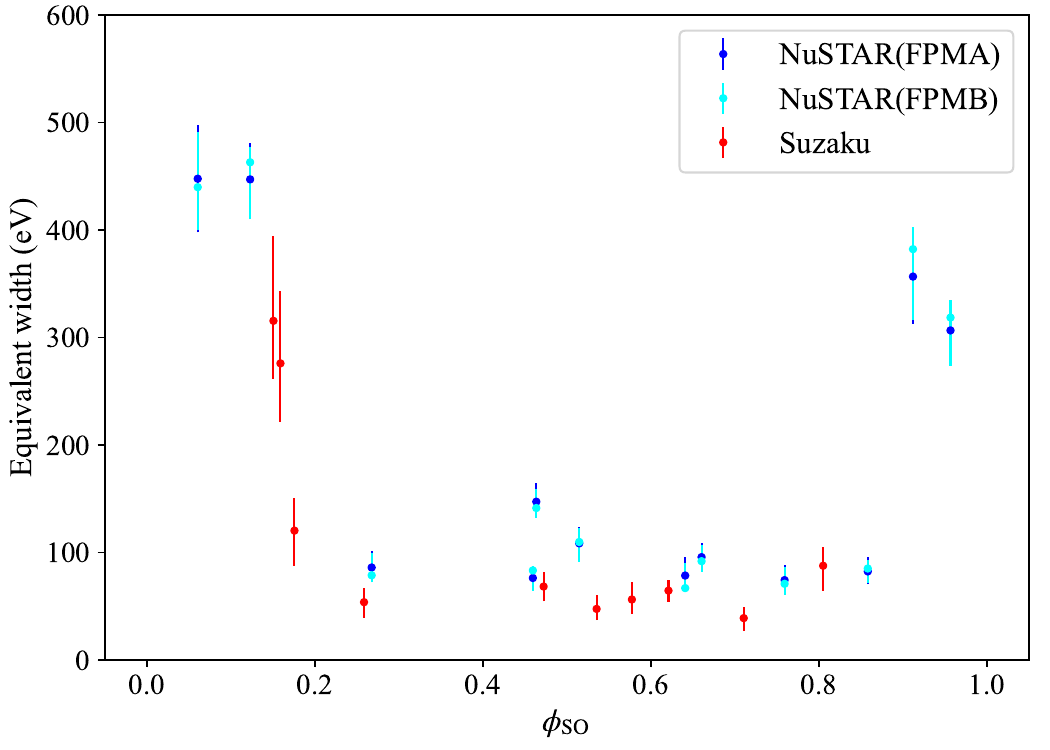}
    \caption{Dependence of the equivalent width of the neutral iron K$\alpha$ line on the superorbital phase ($\phi_\mathrm{SO}$).}
    \label{fig:so_vs_eqwidth}
\end{figure}

By interpolating these derived SO phase values, the superorbital phases corresponding to the Suzaku and NuSTAR observations were determined.
These phases are presented in the second rows of Table \ref{tab:obsSuzaku} and \ref{tab:obsNuSTAR}.
Fig.~\ref{fig:so_vs_nh} shows the variation of the hydrogen column density ($N_\mathrm{H}$) as a function of the superorbital phase for the Suzaku data. 
Observations beginning at early superorbital phases (S40, S60, and S100) show a significant increase in $N_\mathrm{H}$, by nearly an order of magnitude.
Similarly, Fig.~\ref{fig:so_vs_eqwidth} shows a steep decrease in the equivalent widths of the neutral iron K$\alpha$ emission line during the transition from the low state ($\phi_\mathrm{SO}\sim 0$) to the high state.
The interpretation of these $N_\mathrm{H}$ and equivalent width variations will be further elaborated in the subsequent discussion section.

\subsection{Pulse Phase Analysis} \label{sec:timing_analysis_pulse}
We analysed the pulsations of SMC X-1 using data from both Suzaku and NuSTAR observations. 
The energy ranges considered were 15--40\,keV for the Suzaku HXD PIN and 3--40\,keV for NuSTAR.

For precise timing analysis, effects of the Earth's revolution and the neutron star's binary motion must be considered. 
To correct for the former, barycentric correction was applied using \textit{aebarycen} for the Suzaku data. 
For the NuSTAR data, this correction was performed using the \textit{nuproduct} command. 
Subsequently, the Doppler effects arising in the binary motion were corrected by calculating the arrival time at the neutron star's rest frame ($t'$) from the observed time ($t$) using the following equation:
\begin{equation}
    t' = t - \frac{a_x\sin{i}\cos{\phi_\mathrm{orb}}}{c},
\end{equation}
where $\phi_\mathrm{orb}$ represents the orbital phase, $a_x \sin{i}$ is the projected semi-major axis of the neutron star's orbit, and $c$ is the speed of light.
The binary parameters of SMC X-1 used in this data reduction, adopted from \cite{Raichur2010}, are listed in Table \ref{tab:binary_parameters}.
\begin{table}
    \centering
    \caption{Binary motion parameters of SMC X-1}
    \begin{tabular}{ccc}
    \hline
    Parameter & Value & Unit \\
    \hline
    Epoch ($E_0$)             & 42836.1827 & day (MJD)\\
    $a_x \sin{i}$ & 53.5769 & lt--sec\\
    $P_\mathrm{orb}$  & 3.89229263 & day\\
    $\dot{P}_\mathrm{orb}/P_\mathrm{orb}$ & $-3.414\times 10^{-6}$ & $\mathrm{yr}^{-1}$ \\
    \hline
    \multicolumn{3}{l}{These values are referred from \cite{Raichur2010}.}
    \end{tabular} \label{tab:binary_parameters}
\end{table}
Note that the Doppler correction uses not only $a_x \sin{i}$ but also $P_\mathrm{orb}$ and $\dot{P}\mathrm{orb}/P\mathrm{orb}$ at MJD 42836.1827 ($E_0$).
For each observation, we calculated the orbital cycle number since $E_0$ and evaluated the corresponding mid-eclipse time using the updated orbital parameters, including the reduced $\dot{P}\mathrm{orb}$.
We then determined the orbital phase $\phi\mathrm{orb}$ for each photon.
Although \cite{Raichur2010} reported a non-zero eccentricity in a single observation, we assume a circular orbit in this work, consistent with previous studies.

Following these preprocessing steps, we aimed to determine the pulse period $T$ by searching around a candidate value of 0.70\,s. 
For a given trial period $T$, we folded the barycenter- and binary-corrected timing data for both source and background events with a period of $T$. 
A net pulse profile was then obtained by subtracting the background profile from the source profile. 
The counts in each of the $N$ time bins of a single pulse profile are represented as a set $\{F\}_{i=1,..,N}$. 
For each folded pulse profile, we maximized the $\chi^2$ statistic defined as $\chi^2(T;p)=\sum_{i=1}^N \left((F_i - p)/\Delta F_i\right)^2$ by allowing the constant component $p$ to vary. 
This maximization yields the maximum residual $\chi^2_{0}(T)$ for a specific folding period $T$. 
The reduced chi-squared, $\chi^2_{\nu0}$, is then calculated by dividing $\chi^2_0$ by the number of degrees of freedom $\nu = N - 1$.

For each observational dataset, we computed $\chi^2_{\nu0}(T)$ for a range of trial periods $T$. 
In the resulting $T-\chi^2_{\nu0}$ plot, a single Gaussian function was fitted to determine the optimal value of the estimated spin period, which corresponds to the maximum of the $\chi^2_{\nu0}$ distribution. 
Consequently, the pulse period for each observation was determined and is presented in Table \ref{tab:pulsePeriodSuzaku} and Table \ref{tab:pulsePeriodNuSTAR}, with the exception of three observations (S60, N3206, N3712) which occurred during the superorbital low states and for which a coherent pulsation was not clearly detected.
\begin{table}
    \centering
    \caption{Pulse properties of the Suzaku observations}
    \begin{tabular}{ccccc}
        \hline
        ID & 
        $P_\mathrm{spin}$ (s) & $\chi^2_\nu$ & $\mathrm{PF}_1^{\dag}$ & $\mathrm{PF}_2^{\dag}$ \\
        \hline\noalign{\vskip3pt} 
        S10  & 0.701710(4)       & 42.7 & $ 0.48 \pm 0.02$ & $ 0.47 \pm 0.02$\\
        S20  & 0.701699(4)       & 25.8 & $ 0.49 \pm 0.03$ & $ 0.40 \pm 0.03$\\
        S30  & 0.701693(4)       & 25.7 & $ 0.51 \pm 0.03$ & $ 0.38 \pm 0.04$\\
        S40  & 0.70168(2)        & 1.6  & $ 0.6 \pm 0.1$   & $ 0.4 \pm 0.1$\\
        S50  & 0.701635(2)       & 34.6 & $ 0.52 \pm 0.02$ & $ 0.52 \pm 0.02$\\
        S60  & 0.701624$^{\ast}$ & 1.1  & $ 0.24 \pm 0.09$  & $ 0.24\pm 0.09$ \\
        S70  & 0.701530(4)       & 30.8 & $ 0.48 \pm 0.02$ & $ 0.40 \pm 0.03$\\
        S80  & 0.701479(3)       & 33.5 & $ 0.45 \pm 0.02$ & $ 0.44 \pm 0.02$\\
        S90  & 0.701439(5)       & 30.2 & $ 0.56 \pm 0.03$ & $ 0.44 \pm 0.03$\\
        S100 & 0.701346(4)       & 12.6 & $ 0.54 \pm 0.04$ & $ 0.49 \pm 0.05$\\
        \hline 
        \multicolumn{5}{l}{$^\dag$ Errors are determined by statistically. See the main texts for details.}\\
        \multicolumn{5}{l}{$^{\ast}$ Interpolated from the other determined periods.}
    \end{tabular}\label{tab:pulsePeriodSuzaku}
\end{table}
\begin{table}
    \centering
    \caption{Pulse properties of the NuSTAR observations}
    \begin{tabular}{ccccc}
        \hline
        ID & 
            $P_\mathrm{spin}$ (s) & $\chi^2_\nu$ & $\mathrm{PF}_1^{\dag}$ & $\mathrm{PF}_2^{\dag}$ \\
        \hline
        N1001 & 0.701223(3)          & 13.9   & $ 0.18  \pm 0.02 $  & $ 0.08 \pm 0.02 $\\
        N1003 & 0.701184(4)          & 593.2  & $ 0.405 \pm 0.007 $ & $ 0.298\pm 0.008 $\\
        N3202 & 0.699587(3)          & 707.6  & $ 0.364 \pm 0.006 $ & $ 0.359\pm 0.006 $\\
        N3204 & 0.699578(3)          & 467.3  & $ 0.417 \pm 0.007 $ & $ 0.314\pm 0.008 $\\
        N3206 & 0.699536$^{\ast}$    & 1.1    & $ 0.08 \pm 0.02 $  & $ 0.08 \pm 0.03 $\\
        N3208 & 0.699538(3)          & 783.4  & $ 0.386\pm 0.006 $  & $ 0.377\pm 0.006 $\\
        N3702 & 0.697695(6)          & 69.3   & $ 0.31 \pm 0.01 $   & $ 0.17\pm 0.02 $ \\
        N3704 & 0.697680(3)          & 999.3  & $ 0.454\pm 0.006 $  & $ 0.343\pm 0.007 $\\
        N3706 & 0.697671(3)          & 895.9  & $ 0.373\pm 0.005 $  & $ 0.342\pm 0.005 $\\
        N3708 & 0.697335(2)          & 1001.9 & $ 0.363\pm 0.005 $  & $ 0.362\pm 0.005 $\\
        N3710 & 0.697323(3)          & 970.8  & $ 0.475\pm 0.006 $  & $ 0.381\pm 0.007 $\\
        N3712 & 0.697312$^{\ast}$    & 1.0    & $ 0.09 \pm 0.02 $   & $ 0.09 \pm 0.02 $ \\
        \hline
        \multicolumn{5}{l}{$^\dag$ Errors are determined by statistically. See the main texts for details.}\\
        \multicolumn{5}{l}{$^{\ast}$ Interpolated or extrapolated from the other determined spin periods.}
    \end{tabular}\label{tab:pulsePeriodNuSTAR}
\end{table}
\begin{figure*}
  \begin{minipage}[b]{\columnwidth}
    \centering
    \includegraphics[width=\columnwidth]{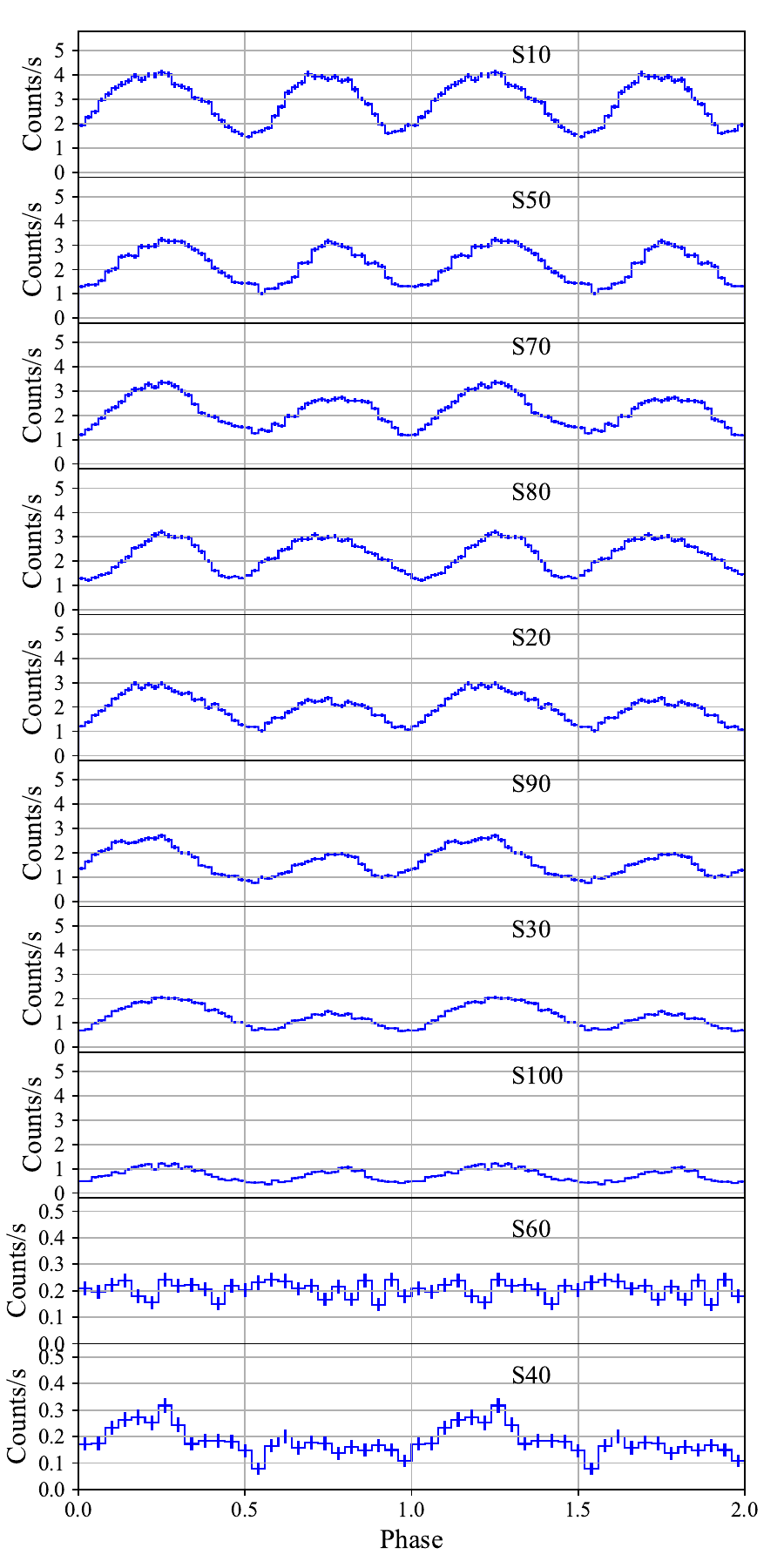}
    \caption{Pulse profiles of Suzaku data. Note that Y-axis ranges of S40/S60 differ from those of other eight observations. The Y-axis errors are $1\sigma$ statistical ones.}
    \label{fig:profile_suzaku}
  \end{minipage}
  \hspace{0.04\columnwidth} 
  \begin{minipage}[b]{\columnwidth}
    \centering
    \includegraphics[width=\columnwidth]{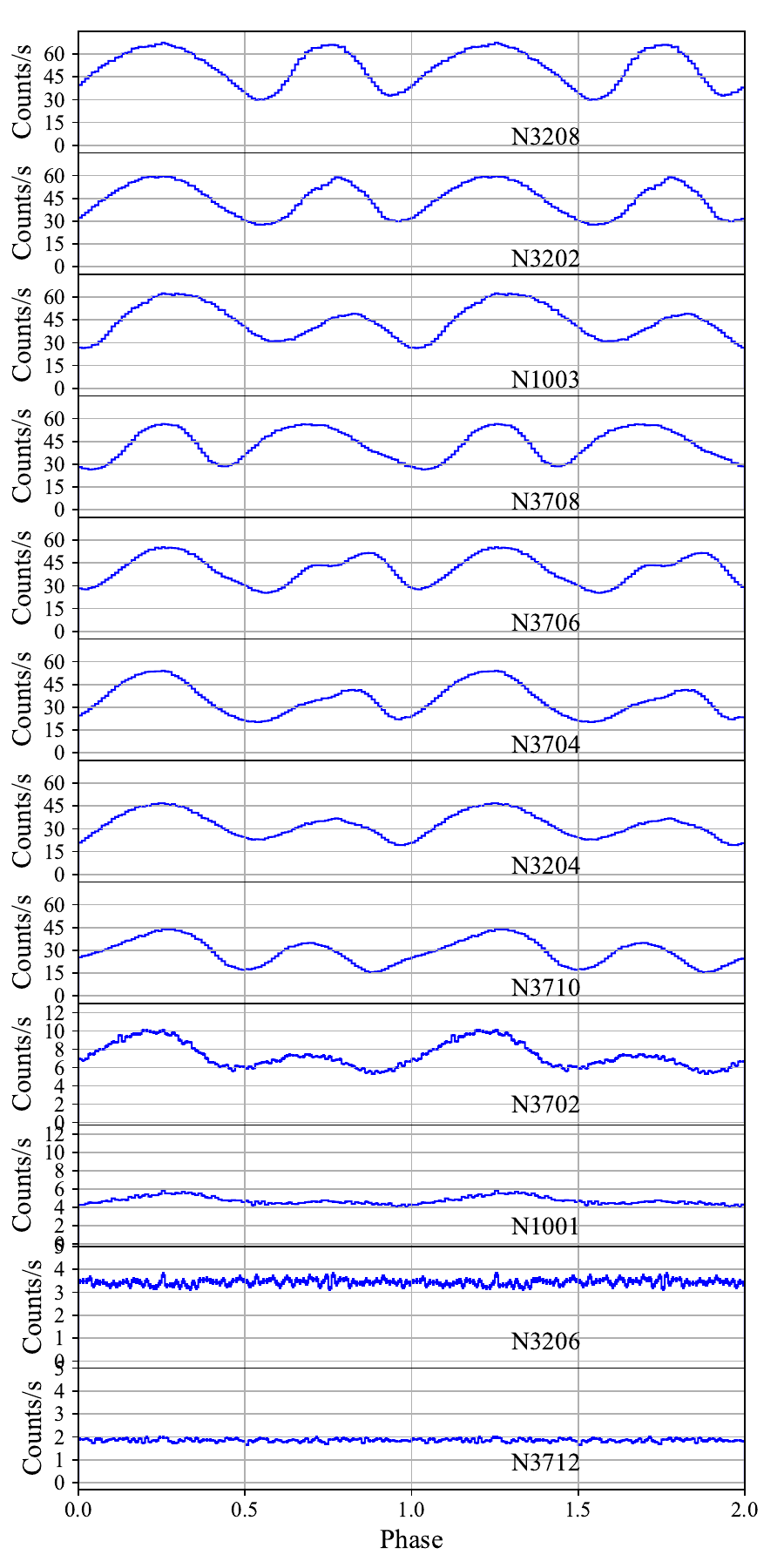}
    \caption{Pulse profiles of NuSTAR data. There are three Y-axis ranges: [0, 12] (N1001), [0, 5] (N3206/N3712), and [0, 75] (the other). The Y-axis errors are $1\sigma$ statistical ones.}
    \label{fig:profile_nustar}
  \end{minipage}
\end{figure*}
Fig.~\ref{fig:profile_suzaku} and Fig.~\ref{fig:profile_nustar} illustrate the pulse profiles derived from all the analysed observations. 
While all observations with detectable pulsations show a double-peaked structure, the relative heights of these peaks vary, potentially correlating with changes in the overall flux.

\begin{figure}
    \centering
    \includegraphics[width=\columnwidth]{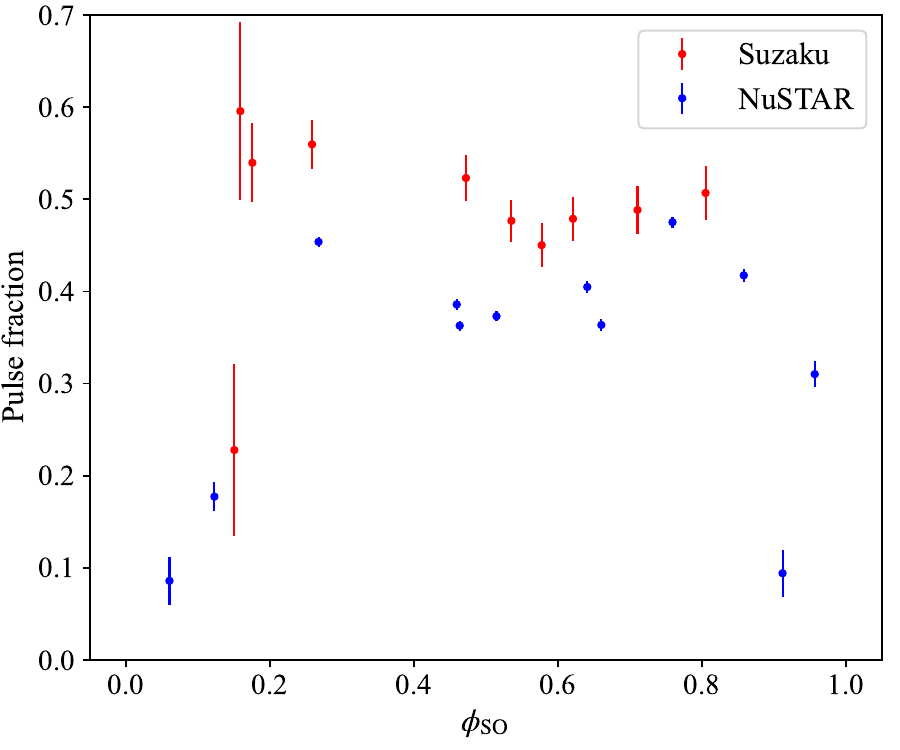} 
    \caption{Dependence of the pulse fraction of the first peak on the superorbital phase.} 
    \label{fig:sophase_vs_PF}
\end{figure}

\begin{figure}
    \centering 
    \includegraphics[width=\columnwidth]{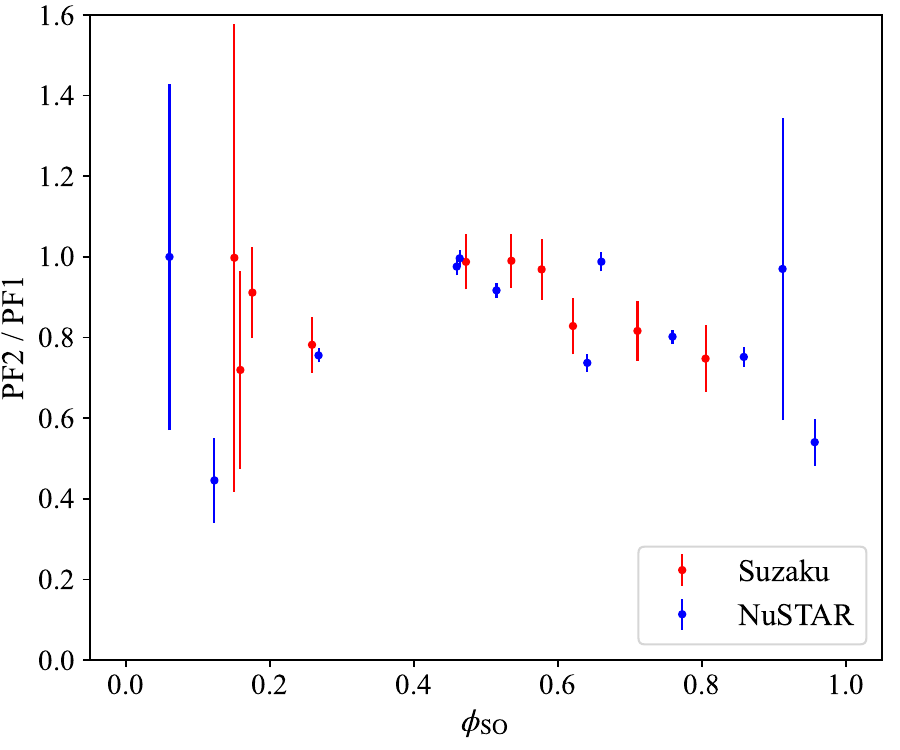} 
    \caption{Dependence of the ratio between the pulse fractions of the first and second peaks on the superorbital phase.}
    \label{fig:sophase_vs_ratioPF} 
\end{figure}

To quantitatively characterize the pulsed emission in each profile, we define the pulse fraction ($\mathrm{PF}_i$) for the $i$-th highest peak ($i=1,2$) as: 
\begin{equation} 
\mathrm{PF}_i = \frac{F_{\mathrm{max},i} - F_\mathrm{min}}{F_{\mathrm{max},i} + F_\mathrm{min}}, 
\end{equation}
where $F_{\mathrm{max},i}$ denotes the flux at the $i$-th peak maximum, and $F_\mathrm{min}$ represents the minimum flux within the corresponding pulse profile. 
By definition, the pulse fractions satisfy $\mathrm{PF}_1 \geq \mathrm{PF}_2$, which typically appears delayed by approximately half a pulse period.
They are listed along with their statistical uncertainties in Table~\ref{tab:pulsePeriodSuzaku} and Table~\ref{tab:pulsePeriodNuSTAR}.
We only consider statistical errors in the calculation of the pulse fractions.
Non-zero pulse fractions, which is not significant, in the three superorbital low phases (S60/N3206/N3712) are caused by neglected systematic errors.
Fig.~\ref{fig:sophase_vs_PF} shows that the pulse fractions of the first peak remain stable during the superorbital high state while they decrease in the low state of the modulation. 
Fig.~\ref{fig:sophase_vs_ratioPF} presents the ratio of the pulse fractions of the first and second peaks, or $\mathrm{PF}_1/\mathrm{PF}_2$.
The second peaks become fainter compared to the first ones in the superorbital low phase.

\section{X-ray transport simulation}
\label{sec:monaco_analysis}
In the previous sections, the superorbital modulation is evident not only in the flux but also in the pulse profiles and in the feature of the neutral Fe K$\alpha$ line.
To investigate the accretion geometry in more detail, we perform photon-transport simulations.

\subsection{Motivation for X-ray transport simulations}
The key features of the X-ray observations are as follows. 
First, the flux in the low state decreases by a factor of $\sim$10 compared with the high state.
Second, in the high state, two pulse peaks of comparable height are present, whereas in the low state no significant peaks are seen (with the exception of S40).
Third, the pulse fraction at the highest peaks exceeds 0.4 in the high state but falls below 0.1 in the low state.
Finally, the equivalent width of the neutral Fe K$\alpha$ line is $EW_{\mathrm{High}}\sim 40\,\mathrm{eV}$ in the high state and exceeds $100\,\mathrm{eV}$ in the low state(s).
In this section, we show that warped and/or tilted accretion-disc geometries can reproduce the observational characteristics found in this work.

Previous studies by \cite{Hickox2005,Brumback2020} performed X-ray transport simulations, assuming a central source with non-uniform bipolar emission and a curved, geometrically thin accretion disc that reflects source photons. 
By associating direct and reflected photons with the hard and soft components, respectively, they partially reproduced features of the hard- and soft-band pulse profiles. 
However, those studies did not address fluorescent line features (e.g., the neutral Fe K$\alpha$ line), spectral dependence, the number of peaks, or the pulse fractions.

We therefore performed photon-transport simulations using the MONACO code \citep{Odaka2011}, which allows arbitrary geometries for both matter and X-ray sources. 
Photons emitted from the neutron star may be reprocessed multiple times via Compton scattering and fluorescence after photoabsorption, before being absorbed or escaping the system. 
The Monte Carlo framework explicitly tracks these multiple interactions in complex geometries, enabling calculations of the observed pulse profiles and spectra as a postprocess.

\subsection{Simulation setup}
\subsubsection{Accretion-disc geometry}
We model the geometry of the accretion system as a combination of a point X-ray source, an optically thick accretion disc, and a disc atmosphere (corona).
The shape of the accretion disc is based on a warped disc model detailed in \cite{Hickox2005}. 
In this model, the disc is expressed by a smooth curved surface formed by two concentric, tilted rings as the inner and outer edges.
The possible origins of superorbital modulation of hard X-rays are not only the optically thick disc, but also the optically thin corona (disc atmosphere) above the disc. 
Both observations and theories of X-ray binaries \citep{Begelman1983a,Begelman1983b,Tomaru2019,Tomaru2023} suggest that such a hot atmosphere is naturally produced by X-ray irradiation from a central source.
\cite{Kubota2018} also suggests the existence of a highly photoionized corona in the 10th observation of Suzaku.

To express the geometry of the curved surface, we use both Cartesian \((x,y,z)\) coordinate and spherical coordinate \((r,\theta,\phi)\).
In addition, angle from the \(xy\)-plane, \(\chi\), is defined to be $\chi=\pi/2 - \theta$.
Relations of those parameters are $x=r\cos{\chi}\cos{\phi}$, $y=r\cos{\chi}\sin{\phi}$, $z=r\sin{\chi}$.
The radii and tilt angles (also measured from the \(xy\)-plane) of the inner and outer edges are denoted by \(r_\mathrm{in}\), \(r_\mathrm{out}\), \(\chi_\mathrm{in}\), and \(\chi_\mathrm{out}\), respectively. 

We introduce a function $\chi_\mathrm{d}(r,\phi)$ to describe the tilt angle of the warped disc surface.
The boundary conditions at the inner and outer edges are $\chi_\mathrm{d}(r_\mathrm{in},\phi)=\chi_\mathrm{in}$, $\chi_\mathrm{d}(r_\mathrm{out},\phi)=\chi_\mathrm{out}$.
A simple representation of the disc surface between $r_\mathrm{in}$ and $r_\mathrm{out}$ is obtained by linearly interpolating the tilt angle with respect to $r$.
To impose point symmetry and allow the surface to remain nearly flat between the most strongly warped directions, a sinusoidal function is multiplied with a phase offset $\phi_\mathrm{tw}$.
Thus, the warped disc surface is expressed as:
\begin{equation}\label{eq:warped_disc}
\chi_\mathrm{d}(r,\phi)
= -\left[ \chi_{\mathrm{in}} + \bigl(\chi_{\mathrm{out}} - \chi_{\mathrm{in}}\bigr)
\frac{r - r_\mathrm{in}}{r_\mathrm{out} - r_\mathrm{in}} \right]\,
\sin\!\bigl(\phi - \phi_\mathrm{tw}\bigr),
\end{equation}
where \(\phi_\mathrm{tw}\) denotes the twist angle between the inner and outer circles. 
Unless otherwise noted, we use the parameters of \cite{Hickox2005}:
\(r_\mathrm{in}=0.8\), \(r_\mathrm{out}=1.0\), \(\phi_\mathrm{tw}=90^\circ\), \(\chi_\mathrm{in}=10^\circ\), and \(\chi_\mathrm{out}=30^\circ\).
Because the model is defined in dimensionless units, we set the physical scale by introducing a reference length \(R_0 = 10^8\,\mathrm{cm}\). 
The physical inner and outer radii are then \(R_\mathrm{in}=R_0\,r_\mathrm{in}\) and \(R_\mathrm{out}=R_0\,r_\mathrm{out}\).

To conduct the Monte Carlo simulation through finite-volume media, we divide the world volume into a spherical \((R,\theta,\phi)\) mesh. 
Each axis has 81 grid points, yielding 80 intervals per direction. 
The volume bounded by adjacent intervals is a \textit{voxel}. 
We classify voxels by location: let the voxel-centre coordinates be \((R_\mathrm{c},\,\theta_\mathrm{c},\,\phi_\mathrm{c})\). 
In addition, \(X_0\) is defined as the adopted half of the angular thickness.
If
\begin{equation}\label{eq:condition_neutral_disc}
\bigl|\chi_\mathrm{c} - \chi_\mathrm{d}(R_\mathrm{c}/R_0,\,\phi_\mathrm{c})\bigr| < X_0,
\end{equation}
the voxel is assigned to the optically thick disc.
Voxels that do not satisfy Equation~\ref{eq:condition_neutral_disc} but do satisfy
\begin{equation}
\left|\chi_\mathrm{c}-\bigl(\chi_\mathrm{d}(R_\mathrm{c}/R_0,\phi_\mathrm{c})\pm\alpha\bigr)\right|<X_0,
\quad \text{with} \quad
\alpha=\arctan\!\left(\frac{H}{r}\right),
\end{equation}
where \(H/r\) is a characteristic scale-height, are classified as corona voxels.
These regions represent fully ionized gas that scatters X-rays from the central neutron star. 
In this work we adopt \(H/r=0.2\) and \(X_0=0.05\,\mathrm{rad}\).
Because our selection on a spherical mesh can leave a gap near the \(\pm Y\) directions (where the warped surface is tangent to the grid shells), we additionally tag the voxels immediately above/below the surface voxels that are exposed to vacuum so that the corona forms a continuous layer.

Number density of the disc material is set to \(n=10^{26}\,\mathrm{cm^{-3}}\), which is so high that incident X-rays are effectively absorbed.
We choose the coronal number density \(n_\mathrm{a}\) so that the scattered flux in the low state is \(\simeq 1/20\) of the high-state flux. 
Let \(\Omega\) be the solid angle subtended by the scattering atmosphere as seen from the X-ray source. Then
\begin{equation}\label{eq:omega_so}
\frac{\Omega}{4\pi}\,\bigl(1-e^{-\tau}\bigr)=\frac{1}{20},
\end{equation}
where \(\tau\) is the Thomson optical depth of the corona. 
For a flat disc with an atmosphere extending by a polar half-angle \(\alpha\), the solid angle \(\Omega\) is expressed as follows:
\begin{equation}\label{eq:omega_geometry}
\Omega=\int_{\pi/2-\alpha}^{\pi/2+\alpha}\!\!\int_0^{2\pi}\sin\theta\,\mathrm{d}\phi\,\mathrm{d}\theta
=4\pi\sin\alpha=\frac{4\pi(H/r)}{\sqrt{1+(H/r)^2}}.
\end{equation}
Combining Equations~\ref{eq:omega_so} and \ref{eq:omega_geometry} with \(H/r=0.2\) gives \(\tau\simeq0.29\). The corresponding density is
\begin{equation}
n_\mathrm{a}=\frac{\tau}{\sigma_\mathrm{T}\,\Delta R}
\simeq \left(4.4\times10^{15}\,\mathrm{cm^{-3}}\right)
\left(\frac{\Delta R}{10^8\,\mathrm{cm}}\right)^{-1},
\end{equation}
where \(\sigma_\mathrm{T}\) is the Thomson cross section and \(\Delta R\) is the characteristic radial thickness of the atmosphere. All simulations in this paper adopt \(\Delta R=10^8\,\mathrm{cm}\).

The iron abundance in the Small Magellanic Cloud (SMC) is taken to be 0.212 relative to the solar value, where the solar abundances are from \citet{Lodders2009} and the SMC abundances are from \citet{Russell1992}.
Gas in each voxel is assumed to corotate with the neutron star at a Keplerian velocity determined by its distance, and this bulk motion is included when computing Doppler shifts and line broadening of interacting photons.

\begin{figure}
    \centering
    \includegraphics[width=0.8\columnwidth]{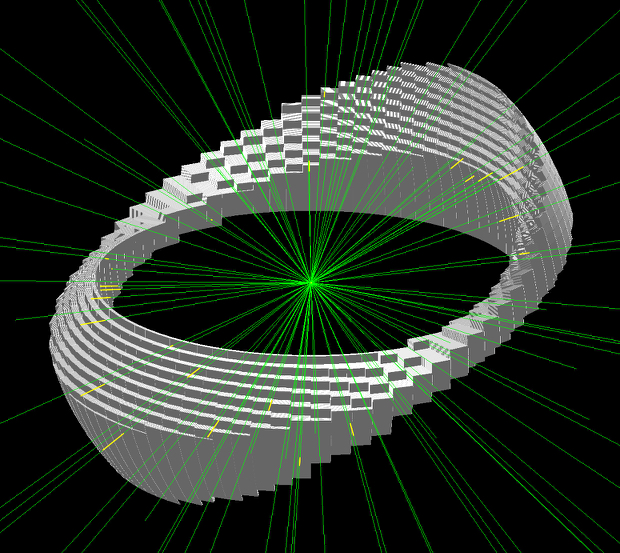}
    \caption{Geometry of the warped-disc model with an isotropic photon source at the origin. White voxels denote the disc and coronal atmosphere; only 100 photon paths are shown for clarity.}
    \label{fig:disc_geometry}
\end{figure}

We simulate \(10^8\) photons with an input spectrum \(\mathrm{d}N/\mathrm{d}E\propto E^{-1}~[{\rm photons~keV^{-1}}]\) over \(1\!-\!100\,\mathrm{keV}\). 
Each photon is classified as a \textit{direct} component if it does not interact with matter during tracking, or a \textit{reflection} component otherwise.
Fig.~\ref{fig:disc_geometry} illustrates the setup: an isotropic point source at the origin illuminates the warped disc and its atmosphere (white voxels). For clarity, only 100 photon trajectories are shown.

\subsubsection{Pulsar emission geometry}
\begin{figure}
    \centering
    \includegraphics[width=\columnwidth]{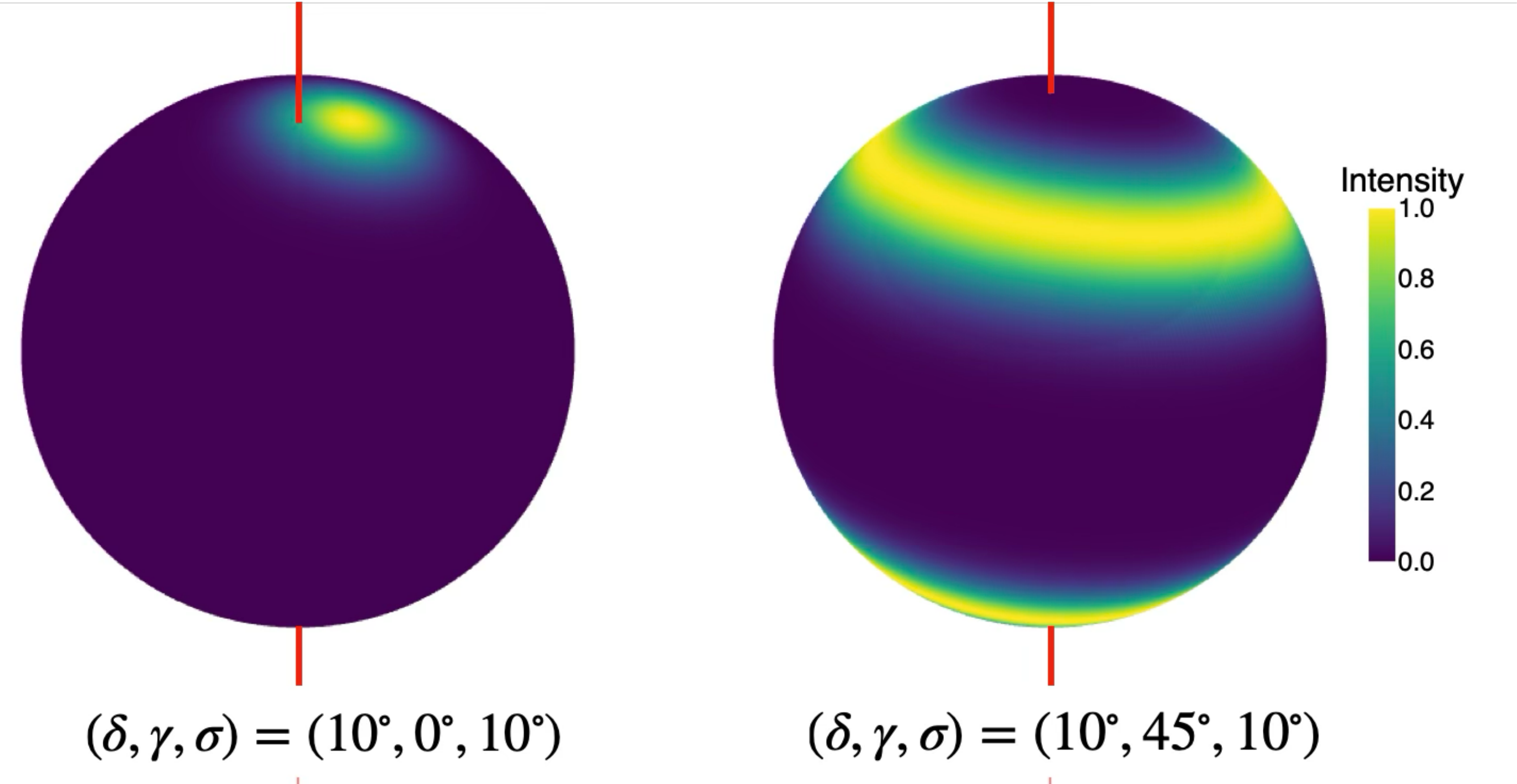}
    \caption{Weight distributions of photon directions on the unit sphere for fixed magnetic poles. The left panel shows emission narrowly aligned with the magnetic pole, while the right panel shows emission broadened by \ang{45}. The red line indicates the spin axis.}
    \label{fig:radProfileEx}
\end{figure}

In our simulation, photons are emitted isotropically from the origin.
Pulsations associated with the accretion columns are imposed a posteriori by assigning each photon with initial direction \(\hat{\boldsymbol{k}}_0\) a weight \(w(\hat{\boldsymbol{k}}_0)\) following \cite{Hickox2005}:
\begin{equation}\label{eq:radModel}
    w(\hat{\boldsymbol{k}}_0)
    = \exp\!\left\{-\frac{(\chi_1-\gamma)^2}{2\sigma^2}\right\}
    + \exp\!\left\{-\frac{(\chi_2-\gamma)^2}{2\sigma^2}\right\},
\end{equation}
where \(\chi_1\) (\(\chi_2\)) is the angle between \(\hat{\boldsymbol{k}}_0\) and the magnetic-axis direction \(\boldsymbol{q}_1\) (\(\boldsymbol{q}_2\)). 
The parameter \(\sigma\) sets the beam width, while \(\gamma\) offsets the peak away from the magnetic axis, or an accretion column.
Fig.~\ref{fig:radProfileEx} shows examples of the directional weights on the unit sphere in velocity space, corresponding to emission narrowly aligned with the accretion column and emission spread by \ang{45} relative to the column.
The red line indicates the spin axis.
In both cases, the magnetic pole is offset by $\delta=10^\circ$ from the spin axis.
The strongest radiation occurs along the direction of the magnetic poles ($\gamma=0^\circ$) in the left pattern, or at an angle of \ang{45} from the poles ($\gamma=45^\circ$) in the right one.
The radiation intensity decreases around the peak following a Gaussian profile with a standard deviation of $\sigma=10^\circ$.

\subsubsection{Observer geometry and sky binning}
An additional parameter that affects the spectra and pulse profiles is the observer’s viewing direction. 
In our setup, occultation by the warped disc depends on the line of sight. 
To sample viewing angles uniformly, we bin photons by sky direction using the Hierarchical Equal Area isoLatitude Pixelization (HEALPix) scheme \citep{gorski2005}.

\begin{figure}
    \centering
    \includegraphics[width=7cm]{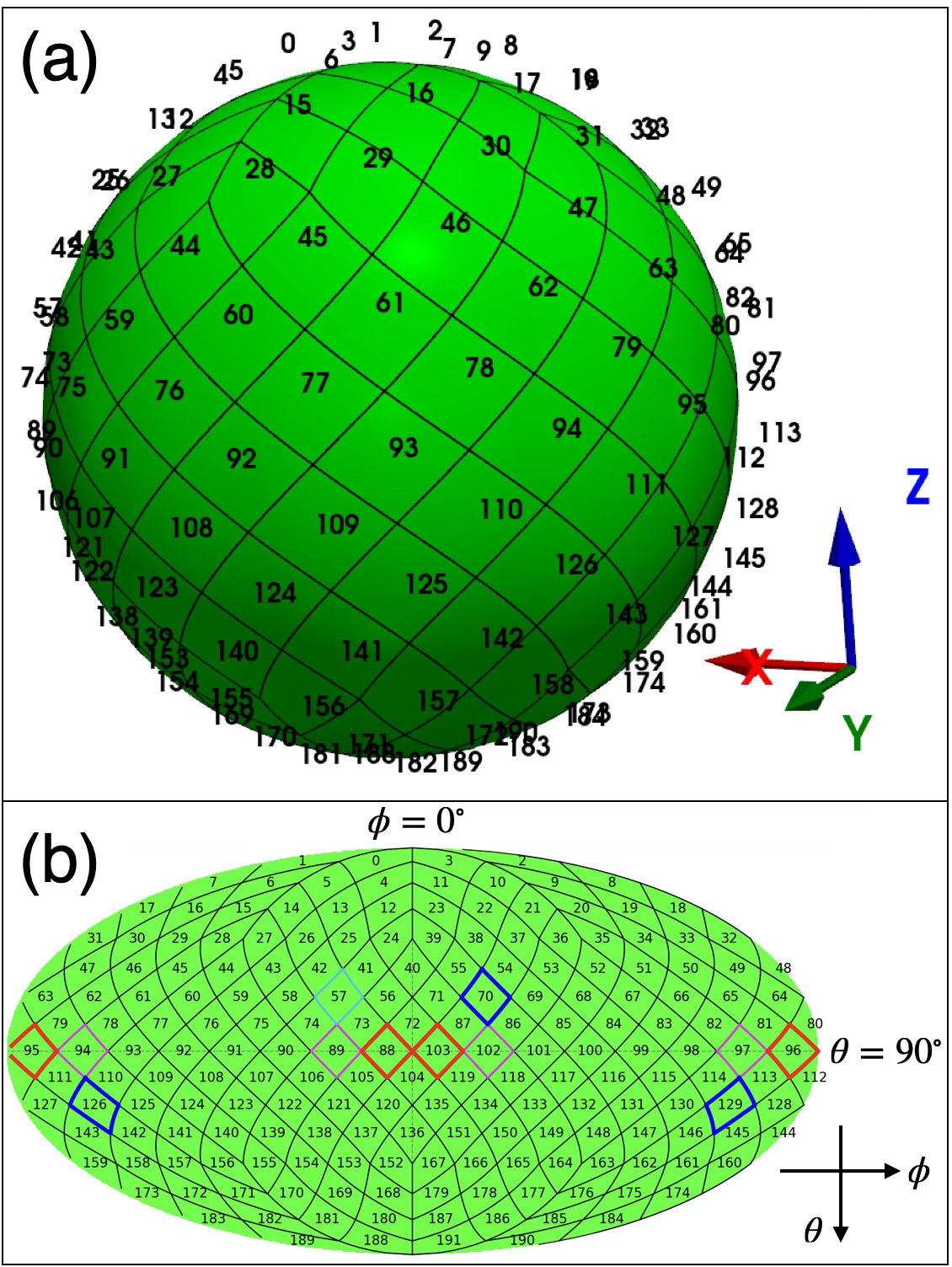}    
    \caption{HEALPix tessellation used to define viewing directions Pixel indices are labeled in the RING ordering scheme. (a) HEALPix pixelization of a sphere at nside = 4, shown in three-dimensional view. (b) The same pixelization is displayed in the Mollweide projection. Coloured pixels are explained in Section \ref{sec:sim_results}.}
    \label{fig:healpix}
\end{figure}

\begin{figure}
    \centering
    \includegraphics[width=\linewidth]{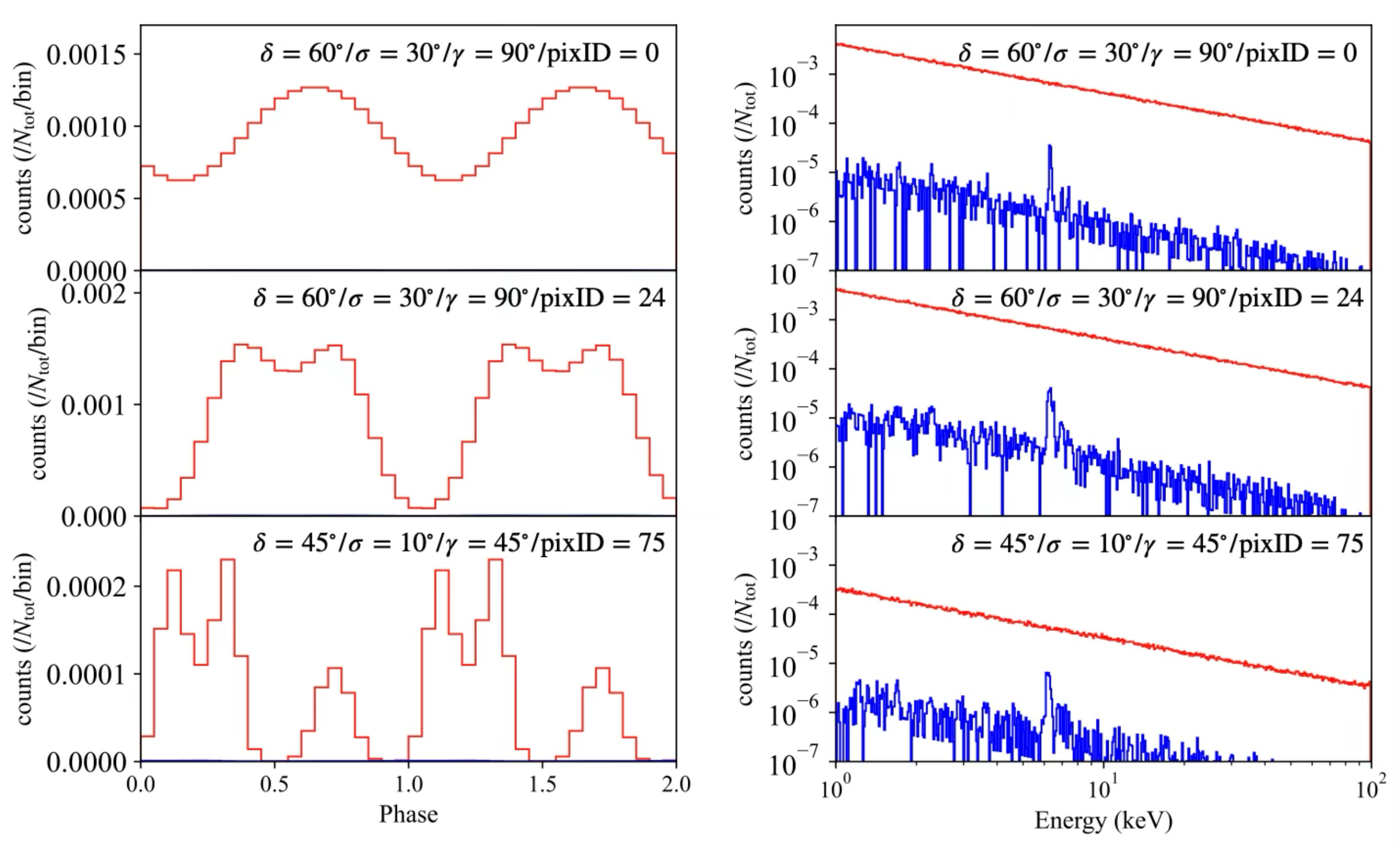}
    \caption{Examples of pulse profiles and spectra. Red and blue lines indicate the direct and reflected components.}
    \label{fig:spectrumProfileDemo}
\end{figure}

We partition the unit sphere into 192 equal-area pixels, which are also represented by $N_\mathrm{side}=4$ in the HEALPix term, as illustrated in Fig.~\ref{fig:healpix}a.
Fig.~\ref{fig:healpix}b shows its Mollweide projection with the corresponding polar angle ($\theta$) and azimuthal angle ($\phi$). 
Each photon is assigned to a single HEALPix pixel corresponding to its outgoing direction, and is accumulated into energy and phase histograms for that pixel.
The contribution of each photon is weighted by \(w(\hat{\boldsymbol{k}}_0)\) from Equation~\ref{eq:radModel}.
Table~\ref{tab:param_summary} summarizes the parameters of the simulation setup. 
The number of combinations of the pulsar emission geometry is \(N_{\delta}\times N_{\sigma}\times N_{\gamma}=4\times3\times6=72\).
For each combination, 192 spectra and pulse profiles, corresponding to the number of HEALPix pixels, are produced.
\begin{table*}
    \caption{Simulation parameters}
    \label{tab:param_summary}
    \begin{tabular}{ccc}
    \hline
    Symbol & Meaning & Value \\ \hline
        \(r_\mathrm{in}\)       & Normalized inner radius & 0.8 \\
        \(r_\mathrm{out}\)      & Normalized outer radius & 1.0 \\
        \(R_0\)                 & Reference length scale (sets physical radii) & \(10^{8}\,\mathrm{cm}\) \\
        \(\chi_\mathrm{in}\)  & Tilt of inner ring (from \(xy\)-plane) & \(10^\circ\) \\
        \(\chi_\mathrm{out}\) & Tilt of outer ring (from \(xy\)-plane) & \(30^\circ\) \\
        \(\phi_\mathrm{tw}\)    & Twist angle between inner and outer rings & \(90^\circ\) \\
        \(\delta \)   & Magnetic-axis colatitude (northern pole) & \((10^\circ,\,30^\circ,\,45^\circ,\,60^\circ)\) \\
        \(\gamma\) & Fan-beam offset angle in Eq.~\ref{eq:radModel} & \((0^\circ,\,10^\circ,\,45^\circ,\,60^\circ,\,75^\circ,\,90^\circ)\) \\ 
        \(\sigma\)   & Beam width parameter \(\sigma\) in Eq.~\ref{eq:radModel} & \((10^\circ,\,30^\circ,\,45^\circ)\) \\
        \hline
    \end{tabular}
\end{table*}
As examples of the calculated pulse profiles and spectra normalized by the number of the initial photons ($N_\mathrm{tot}=10^8$), three sets are shown in Fig.~\ref{fig:spectrumProfileDemo}.
The red and blue lines indicate the direct and reflected components.
The top and middle rows correspond to \((\delta,\,\sigma,\,\gamma)=(60^\circ,\,30^\circ,\,90^\circ)\) with different viewing directions (HEALPix pixel ID).
In the middle panels, double peaks appear, but the flux between the peaks does not return to the baseline. This behaviour is not seen in the high state of SMC X-1.
The bottom row shows the case \((\delta,\,\sigma,\,\gamma)=(45^\circ,\,10^\circ,\,45^\circ)\).
Here, radiation from one magnetic pole is attenuated, and radiation from the other pole is blocked by the optically thick disc around phase \(\approx 0.25\). 
As a result, three peaks appear.
Although a \SI{6.4}{keV} neutral iron line is present in the reflected component, it can be easily hidden by the direct component.
To increase the equivalent width of the iron line, the accretion disc must subtend a larger solid angle to the observer (for that HEALPix pixel).

\subsection{Evaluation Criteria for Simulation Results}
The goal of our simulation is to test whether precession of the warped disc can explain the changes in (i) the pulse fraction of the first peak and (ii) the equivalent width (EW) of the neutral K$\alpha$ iron line.
In the simulation, the neutron star and the disc are fixed; changing the observer’s direction mimics the superorbital phase.
The superorbital modulation is expressed by the difference of HEALPix IDs.
For every combination of $\delta$, $\sigma$, $\gamma$, and HEALPix pixel ID (Table~\ref{tab:param_summary}), we calculate the EW of the \SI{6.4}{keV} line and several pulse fractions.
We exclude polar pixels with polar angle $\theta<30^\circ$ or $\theta>150^\circ$, because the disc is unlikely to precess across the line of sight there.
For convenience, we define several quantities as follows.
Let $F_\mathrm{max}$ and $F_\mathrm{min}$ be the maximum and minimum values of the pulse profile, respectively.
We further define $F_\mathrm{val}$ as the minimum flux in the valley between the two strongest peaks, excluding the valley that contains the global minimum $F_\mathrm{min}$.
"High (low) state" means a viewing direction with large (small) phase-averaged 15--40\,keV flux.

Thus, the simulation should reproduce the following features:
\begin{enumerate}
    \item In the high state, the second-highest peak has $\mathrm{PF}>0.4$, a third peak (if present) has $\mathrm{PF}<0.1$, and no peak exceeds $\mathrm{PF}=0.7$.
    \item In the high state, the two main peaks are separated by a deep valley. We require
    \begin{equation}
        \frac{F_\mathrm{val}-F_\mathrm{min}}{F_\mathrm{max}-F_\mathrm{min}} < 0.1,        
    \end{equation}
    i.e., the inter-peak flux is close to the baseline.
    \item The phase-averaged 15--\SI{40}{keV} flux ratio between low and high states satisfies
    \begin{equation}
    0.002 \le \frac{F_{15-40\,\mathrm{keV}}^{\mathrm{(low)}}}{F_{15-40\,\mathrm{keV}}^{\mathrm{(high)}}} \le 0.2 .   
    \end{equation}
    \item In the low state, the Fe K$\alpha$ EW at \SI{6.4}{keV} is $\ge \SI{100}{eV}$.
\end{enumerate}

\subsection{Results}\label{sec:sim_results}
\begin{figure}
    \centering
    \includegraphics[width=0.6\linewidth]{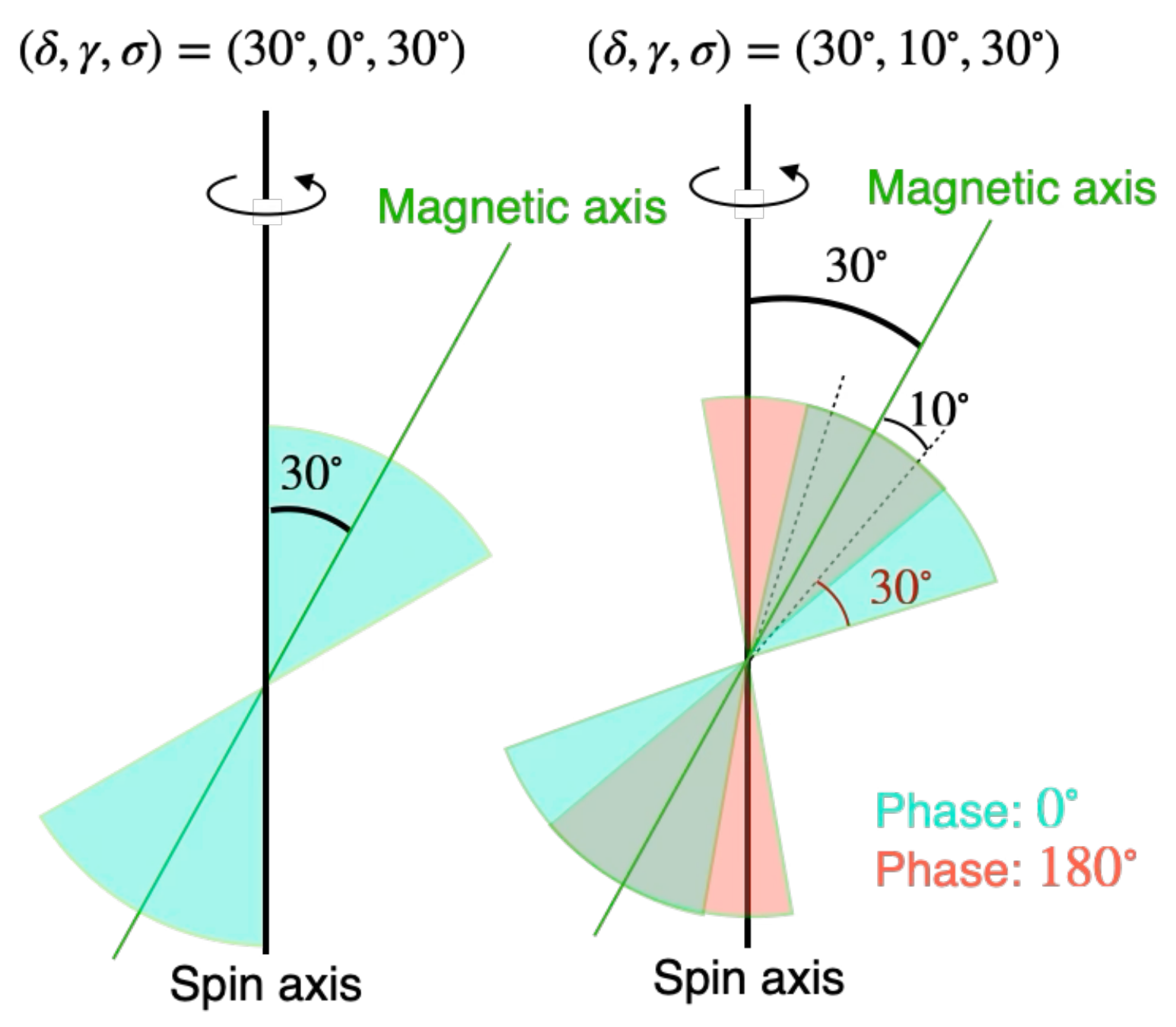}
    \caption{Instantaneous emission directions weighted by Equation~\ref{eq:radModel}. The left and right panels correspond to the first and second rows in Table~\ref{tab:par_survive}, respectively. In the right panel, emission profiles for two rotational phases ($0^\circ,\ 180^\circ$) are shown in cyan and red, respectively.}
    \label{fig:radPattern}
\end{figure}

The number of selected combinations is severely limited for observer direction, radiation direction and its spread, as shown in Table \ref{tab:par_survive}.
Except for a small misalignment between the emission centre and the magnetic pole ($\gamma$), the results favour a misalignment angle between the spin axis and the magnetic pole of $\delta=30^\circ$ with an emission spread angle of $\sigma=30^\circ$ (left panel of Fig.~\ref{fig:radPattern}).
Among 72 combinations of the radiation patterns characterized by $\delta,\ \sigma,\ \gamma$, only two cases can meet the evaluation criteria.
For $\gamma=0^\circ$, or case 1, HEALPix IDs 88, 95, 96, and 103 satisfy all the conditions in the high state (red in Fig.~\ref{fig:healpix}b), whereas IDs 70, 126, and 129 do so in the low state (blue in Fig.~\ref{fig:healpix}b).
For $\gamma=10^\circ$ (case 2), in addition to these pixels, HEALPix IDs 89, 94, 97, 102 also satisfy the high state conditions (magenta in Fig.~\ref{fig:healpix}b), and pixel 57 does so in the low state (cyan in Fig.~\ref{fig:healpix}b).

\begin{figure}
    \centering
    \includegraphics[width=\columnwidth]{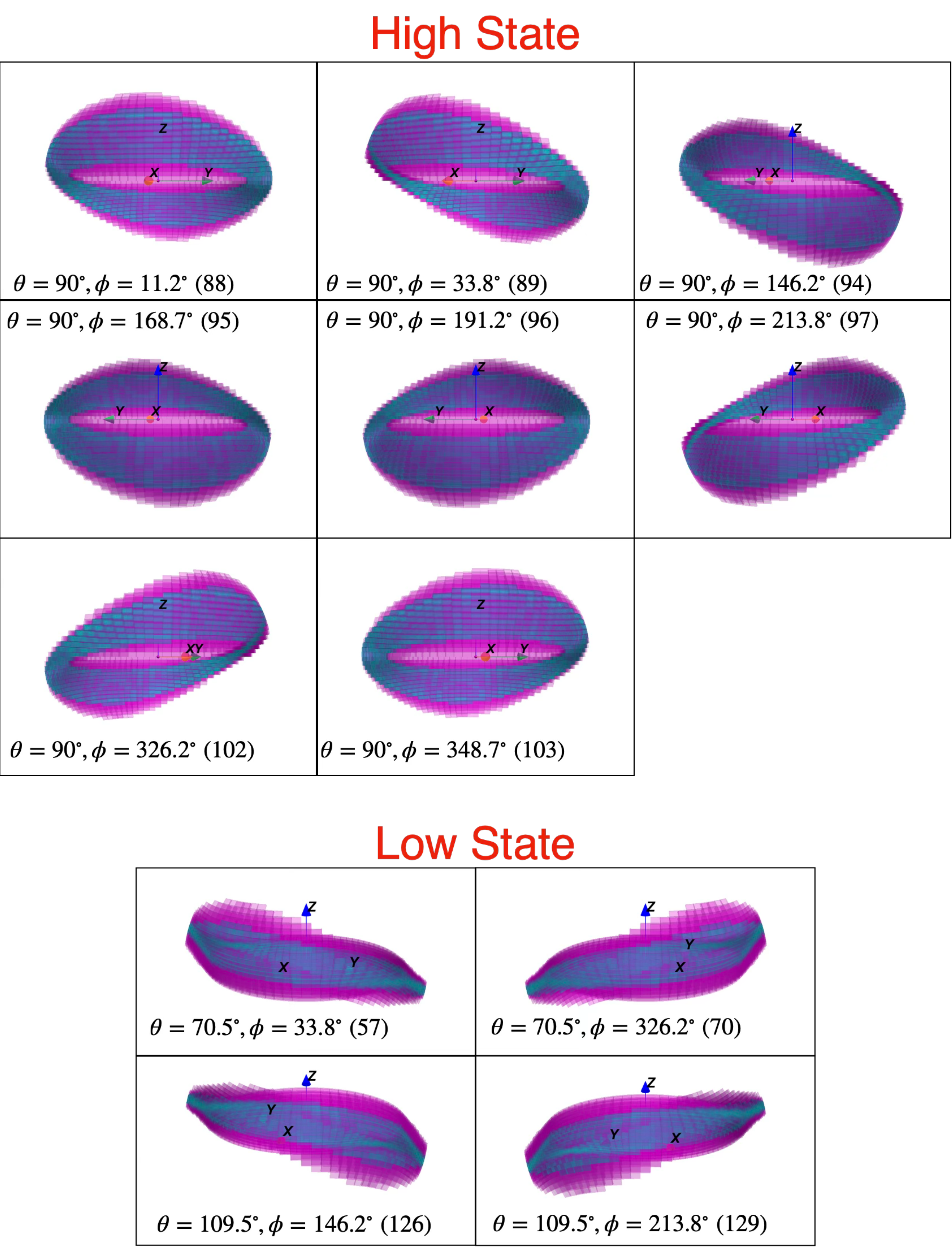}
    \caption{The simulated accretion system viewed from directions corresponding to HEALPix ID listed in Table \ref{tab:par_survive}. The optically thick discs and the corona are shown as cyan and magenta blocks, respectively.}.
    \label{fig:disc_pix_viewing}
\end{figure}

Fig.~\ref{fig:disc_pix_viewing} shows schematics of the accretion system as viewed from the HEALPix directions for all allowed conditions listed in Table~\ref{tab:par_survive} for the high state (the top) and the low state (bottom).
\begin{table*}
    \caption{Parameters satisfying the condition described in the text}
    \label{tab:par_survive}
    \begin{tabular}{ccccccc}
    \hline
    Case & $\delta$ & $\gamma$ & $\sigma$  & \multicolumn{2}{c}{Pixel ID$^\ast$} & \# of Patterns\\
    & & & & High & Low & \\
    \hline
    1 & $30^\circ$ & $0^\circ$ & $30^\circ$ & (88, 95, 96, 103) & (70, 126, 129) &12\\
    2 & $30^\circ$ & $10^\circ$ & $30^\circ$ & (88, 89, 94, 95, 96, 97, 102, 103) & (57, 70, 126, 129) & 32\\
    \hline
    \multicolumn{7}{l}{$^\ast$ Any combination of one HEALPix pixel ID from the high state and one from the low state satisfies the observational characteristics.}
    \end{tabular}
\end{table*}
The line of sight in each panel is specified by the polar ($\theta$) and azimuthal ($\phi$) angles, with the corresponding HEALPix pixel number (ring ordering) given in parentheses. 
An accretion disc (cyan) is surrounded by corona blocks (magenta), and the central X-ray source is located at the origin.
During the high state, the central star is directly visible to the observer; by contrast, in the low state, the accretion disc blocks the direct photons from the source.

\begin{figure*}
    \centering
    \includegraphics[width=16cm]{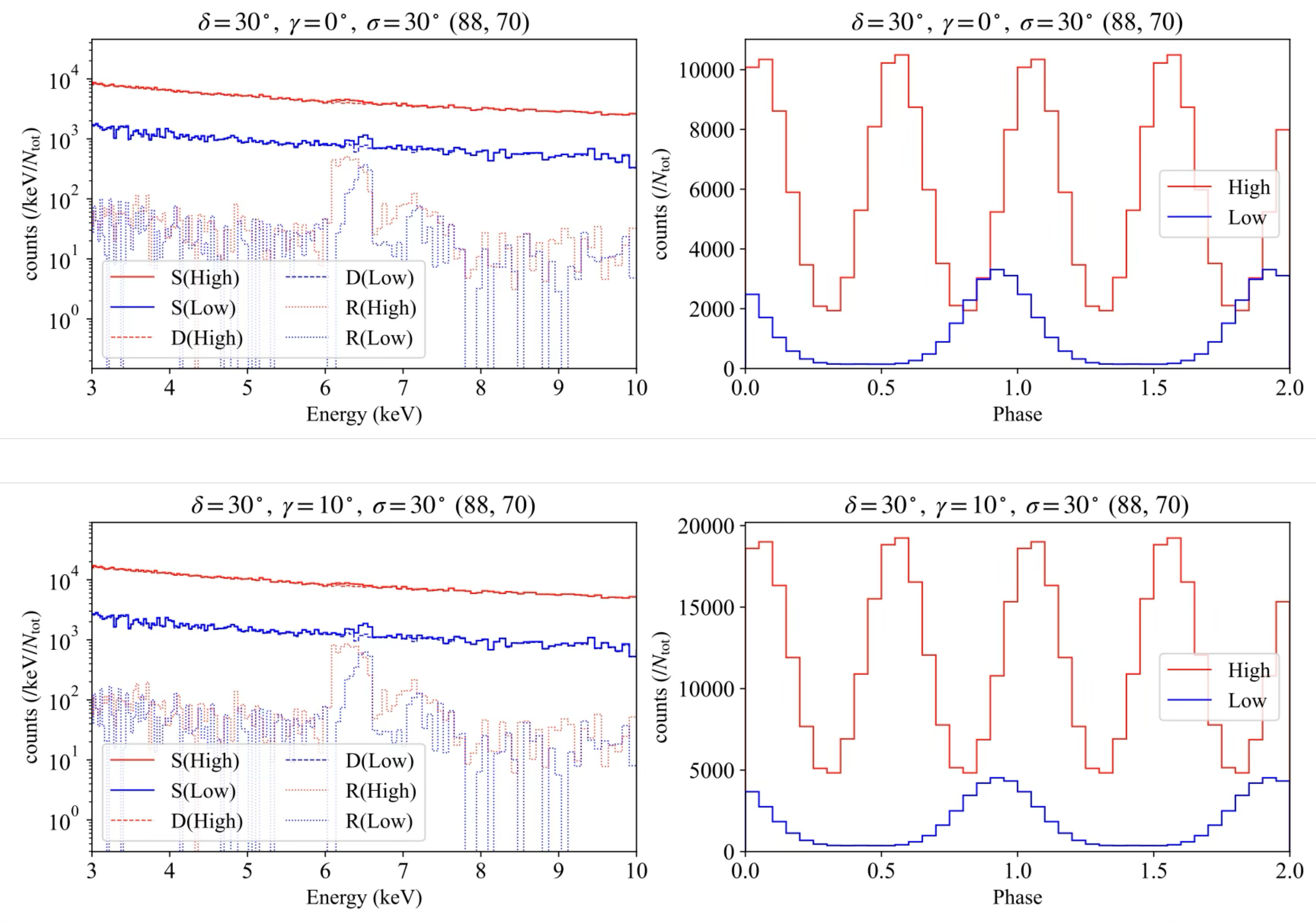}
    \caption{Simulated spectra and pulse profiles for the cases in Table \ref{tab:par_survive} with the HEALPix pixel IDs of 88 (high state) and 70 (low state). The top two panels and the bottom two panels correspond to the case 1 $(\delta=30^\circ,\gamma=0^\circ,\sigma=30^\circ)$ and the case 2 $(\delta=30^\circ,\gamma=10^\circ,\sigma=30^\circ)$, respectively.}
    \label{fig:spectra_profiles_survived}
\end{figure*}

Fig.~\ref{fig:spectra_profiles_survived} shows examples of the high and low states for case 1 and case 2, as listed in Table \ref{tab:par_survive}.
In both spectra and pulse profiles, red lines (blue) indicate high (low) state results.
The dashed and dotted lines are reflection and direct components, respectively.
While the peak flux of the iron line for both cases does not show significant variation, the direct component buries that feature, which results in large equivalent width in low states.
In our simulation framework, Keplerian velocities are assigned to the accreting materials, resulting in rotational motions around the $z$-axis. 
Doppler effects are taken into account during photon interactions in the transport calculation. 
As a result, the neutral iron line produced in the reflected component is blue-shifted when emission from material moving toward the observer dominates, and red-shifted otherwise.
In the pulse profiles, two distinct peaks appear in high state.
Features in the spectra and pulse profiles are similar among the other 43 patterns.
Among the HEALPix pixel combinations, the three pairs (High, Low) = (95, 126), (96, 129), and (103, 70) correspond to an accretion-disc precession of only \ang{29.4}.
This means approximately difference of \ang{30} precession can reproduce the detailed evaluation criteria.

\section{Discussion} \label{sec:discussion}
In this section, we discuss the origin of the superorbital modulation of SMC X-1 and its accretion flow geometry.
We consider the scenario of obscuration by a warped accretion disc, and the possibility of the intrinsic luminosity variation.
Our discussion is based on the results of spectral and timing analyses (Section \ref{sec:spectral_analysis}, \ref{sec:timing_analysis}) and the simulation (Section \ref{sec:monaco_analysis}).

First, if the intrinsic luminosity variation of the neutron star were the dominant cause of the superorbital modulation, both the continuum and the reprocessed line emission would vary in the same manner.
In such a case, no significant variation of the equivalent width would be expected. 
However, the equivalent width is significantly enhanced in the superorbital low state ($\phi_\mathrm{SO}<0.2$ or $\phi_\mathrm{SO}>0.9$), as shown in Fig.~\ref{fig:so_vs_eqwidth}.
In addition, Figure~\ref{fig:so_vs_nh} shows a sharp increase in $N_\mathrm{H}$ during the superorbital low state, while it remains stable in the high state.
These observational facts support that the decrease in the X-ray luminosity is attributed to attenuation by matter, which is plausibly associated with the warped accretion disc.

The iron line spectroscopy provides information on the structure of the accretion material. 
Our analysis shows that the neutral iron emission line has a broad width of $\sim$ \SI{0.3}{keV} in the Suzaku data (Table \ref{tab:fit_param_suzaku}). 
If Doppler broadening is responsible for this width at the $\mathrm{K\alpha}$ energy of \SI{6.4}{keV}, the velocity of the material is $\sim 1.4\times10^4\,\mathrm{km\,s^{-1}}$. 
Assuming Keplerian motion around a $1.06\,M_\odot$ neutron star, this velocity corresponds to an orbital radius of \SI{7e7}{cm}. 
This value is consistent with the Alfv\'{e}n radius of $\sim$ \SI{1.3e8}{cm} (see Appendix. \ref{sec:alfven_radius}), within a factor of two. 
Although neutral iron line emission can arise from the stellar wind of the donor star, the observed velocity of the line reprocessing matter is even faster than a typical wind velocity of a B0-type star, which is less than \SI{2000}{km.s^{-1}} \citep{Bernabeu1989}.
Therefore, the iron line originates from the reflection at the inner accretion disc that truncates around the Alfv\'{e}n radius.

In the spectral analysis of SMC X-1, the apparent variation of the continuum over the superorbital modulation is most likely explained by changes in the low-energy absorption and by variations in the relative strength of the Fe $\mathrm{K\alpha}$ line, rather than by intrinsic changes in the continuum itself.
In the Suzaku data, stronger absorption in the S40 and S60 phases suppresses the low-energy flux, which can result in higher cutoff energies $E_\mathrm{c}$.
This implies that the spectral shape vary significantly, even though a mildly negative correlation between $N_\mathrm{H}$ and $E_\mathrm{c}$ is seen within individual observations (see Appendix. \ref{sec:corner_plot}).
The decrease in the folding energy from $\sim$ \SI{8}{keV} in the high state to $\sim$ \SI{5}{keV} in the low state is likely related to the enhanced \SI{6.4}{keV} Fe line (with a larger equivalent width), whose reduction affects the fitted cutoff shape.
Similarly, in the NuSTAR data, increased absorption during the superorbital low state may shift the NPEX power-law index $\alpha$ toward smaller values (0.05–0.3 compared with 0.39–0.45 in the high state), resulting in a harder continuum.

The variation in the pulse shape can be attributed to the blocking effect of the accretion disc.
Fig.~\ref{fig:sophase_vs_PF} shows high pulse fractions in the high state; the pulse fraction decreases with the luminosity, and the pulsation almost disappears in the low state.
Thus, the neutron star pulsation is smeared out by the warped disc in the low state.
Furthermore, Fig.~\ref{fig:sophase_vs_ratioPF} presents a relative decrease of the second peak pulse fraction compared to the first one.
This suggests that emission from the antipodal magnetic pole is more effectively hidden by the disc during the low states.
Although the intrinsic change in the accretion columns on the magnetic poles could be responsible for the variation of the pulse profile, the superorbital variability in the pulsation is naturally explained by the partial covering of the neutron star by the inner region of the warped disc.

We consider the simulation results described in Section \ref{sec:monaco_analysis} in the context of the warped disc interpretation.
We found that partial occultation by an optically thick disc can reproduce the observed superorbital variations in the pulse shape, the iron line strength, and the total flux.
As shown in Fig.~\ref{fig:spectra_profiles_survived}, the reflection continuum, which includes scattered component by the corona, is much fainter than the direct continuum from the neutron star, even in the low-state spectrum.
The simulation parameter sets that well reproduce the observation strongly constrain the emission geometry.
The neutron star emission comes from two symmetrically located magnetic poles, the angular profile of the emission from each pole is mildly spread, $\sigma\sim 30^\circ$, and the beam offset angle $\gamma$ is small, $\lesssim10^\circ$.
A disc precession angle of \ang{30} reproduces the observed modulation.
\cite{Forsblom2024} analysed observations of SMC X-1 by the Imaging X-ray Polarimetry Explorer (IXPE) in December 2023.
They fitted the phase-resolved variations of the polarisation angle with a model function based on the rotating vector model \citep{Radhakrishnan1969,Meszaros1988,Poutanen2020}.
The misalignment angles between the spin and magnetic axes (Table 7 in that paper) are consistent, in most cases, with our result ($\delta = 30^\circ$) within the $2\sigma$ uncertainties.

Finally, we summarize the possible scenario.
In the superorbital low state, the direct X-rays are blocked by the inner region of the warped accretion disc, while the reflection by the disc including the iron line remains. 
As a result, the equivalent width of the iron K$\alpha$ line increases as the continuum decreases.
This obscuration naturally explains the increase of $N_\mathrm{H}$ in the low state.
The gradual decrease of the X-ray flux in the superorbital modulation is possibly explained by the partial covering of the accretion column emission by relatively small disc edge/surface structure.
Our simulation successfully reproduces the superorbital variability of SMC X-1, well fitted to the precessing disc attenuation scenario.
This model does not require the intrinsic change of the accretion column to explain the observation.
The stability of the accretion of SMC X-1 is consistent with a commonly accepted picture of accretion-powered pulsars, in which a disc-fed, high-luminosity accreting pulsar are stable, but wind-fed, low-luminosity pulsars are highly variable \citep{Mushtukov2022, Tamba2023, Tamba2025, Odaka2014}.

There are several effects that are not fully taken into account in our simulation.
In the model, the superorbital modulation is represented by changes in the HEALPix IDs, with the relative configuration of the neutron star, disc, and corona kept fixed.
This simplified treatment effectively alters the inclination angle of the neutron star during the superorbital modulation.
Another possible issue arises from the finite solid angle of a HEALPix pixel within which all photons are integrated to extract a spectrum and a pulse profile for one simulation parameter set.
In reality, the observing pixel should be infinitely small, but we reinterpret this as a partial covering effect of the finite size of the neutron star emission in this work.
The obscuring matter is possibly composed of small clumps or inhomogeneous surface structure of the accretion disc.
The existence of the inhomogeneous X-ray absorbing clumps is consistent with the high $N_\mathrm{H}$ value observed in the superorbital low state.
Note that our simulation approach could reproduce a wider variety of spectra and temporal features, ranging from superorbital low and transition states to high states. 
However, the time variability seen in detailed observations strongly depends on the spatial distributions of (non-)ionized matters around the neutron star, which requires more comprehensive understanding of the accretion system. 
A detailed treatment of these effects in the Monte Carlo simulation is beyond the scope of this work.

\section{Conclusions} \label{sec:conclusions}
We analysed all available Suzaku and NuSTAR X-ray observations of SMC X-1 taken between 2011 and 2022 to investigate its accretion structure, focusing on the quasi-periodic superorbital modulation. 
In the spectral analysis, the Suzaku and NuSTAR spectra in the 3--40~\si{keV} range were fitted with a continuum model plus a Gaussian component at 6.4\,keV which represents the neutral iron K$\alpha$ emission line. 
The hydrogen column density ($N_\mathrm{H}$) was estimated from the Suzaku spectra. 
Two low-state observations (S40 and S60) showed significantly higher $N_\mathrm{H}$ values ($\sim 10^{23}\,\mathrm{cm}^{-2}$), compared with $3 \times 10^{22}\,\mathrm{cm}^{-2}$ in the intermediate state (S100) and $\sim 1 \times 10^{22}\,\mathrm{cm}^{-2}$ in the brighter states. 
The superorbital phase was determined using the Hilbert-Huang transform, which enables time-resolved phase assignment from MAXI monitoring data. 
By correlating these phases with the spectral results, we found that the equivalent width of the neutral iron K$\alpha$ line increases markedly during the superorbital low states.
Furthermore, the pulse fractions of the first peaks in the pulse profiles remain stable during the superorbital high state while they decrease in the low state of the modulation.
The second peaks are less prominent than the first ones during the superorbital low phase.

By combining the observational results with photon-transport simulations, the superorbital modulation observed in SMC X-1 can be explained by X-ray attenuation caused by a precessing, warped accretion disc.
Occultation of the central source by the precessing disc successfully reproduces the observed variations in the iron K$\alpha$ equivalent width, the pulse profiles, and the hard-X-ray continuum flux.
Notably, a disc precession angle of approximately \ang{30} can account for the observed features. 
For the radiation pattern of the photon source, the preferred beam width corresponds to a standard deviation of \ang{30}.
We found no evidence for intrinsic variability in the luminosity of the neutron star.

\section*{Acknowledgements}
The authors thank an anonymous reviewer for insightful comments which improve the manuscript.
This work has been supported by KAKENHI Grant-in-Aid for Scientific Research on Innovative Areas No. 18H05861, 19H01906, 19H05185, 22H00128, 22J11653, 22K18277, 22KJ0811, 23K13147, 23K25907, 24KJ0152, and by Toray Science and Technology Grant No. 20-6104 (Toray Science Foundation). 
ST was supported by RIKEN Junior Research Associate Program.
Supercomputer HOKUSAI Bigwaterfall2 system was used for the Monte Carlo simulation as a RIKEN project (RB240053).

\section*{Data Availability}

The Suzaku and NuSTAR observations are publicly available through HEASARC (\url{https://heasarc.gsfc.nasa.gov/docs/archive.html}) while MAXI data can be downloaded from its web site (\url{https://wmaxi.riken.jp/top/index.html}).



\bibliographystyle{mnras}
\bibliography{reference}



\appendix
\section{Alfv\'{e}n radius of SMC X-1}
\label{sec:alfven_radius}
The Alfv\'{e}n radius of a spherical accretion system, $r_\mathrm{A}$, can be derived as \citep{Lamb1973,Davidson1973}:
\begin{equation}
    r_\mathrm{A} \simeq (GM)^{-1/7}\mu^{4/7} \dot{M}^{-2/7},
\end{equation}
where $M$ is the neutron star mass and $G$ is the gravitational constant.
For SMC X-1, the neutron star mass is $1.06M_{\odot}$ \citep{VanderMeer2007}.
By substituting $L=5\times 10^{38}~\mathrm{erg\,s^{-1}}$ and $B=10^{12}~\mathrm{G}$ into the above equation, the Alfv\'{e}n radius is estimated as

\begin{equation}
	r_\mathrm{A} \simeq G^{1/7}M^{1/7}R^{10/7}B^{4/7}L^{-2/7} \sim 1.3\times 10^{8}~\mathrm{cm} .
\end{equation}

\section{Corner plots for Suzaku data}
\label{sec:corner_plot}
\begin{figure}
    \centering
    \includegraphics[width=0.9\linewidth]{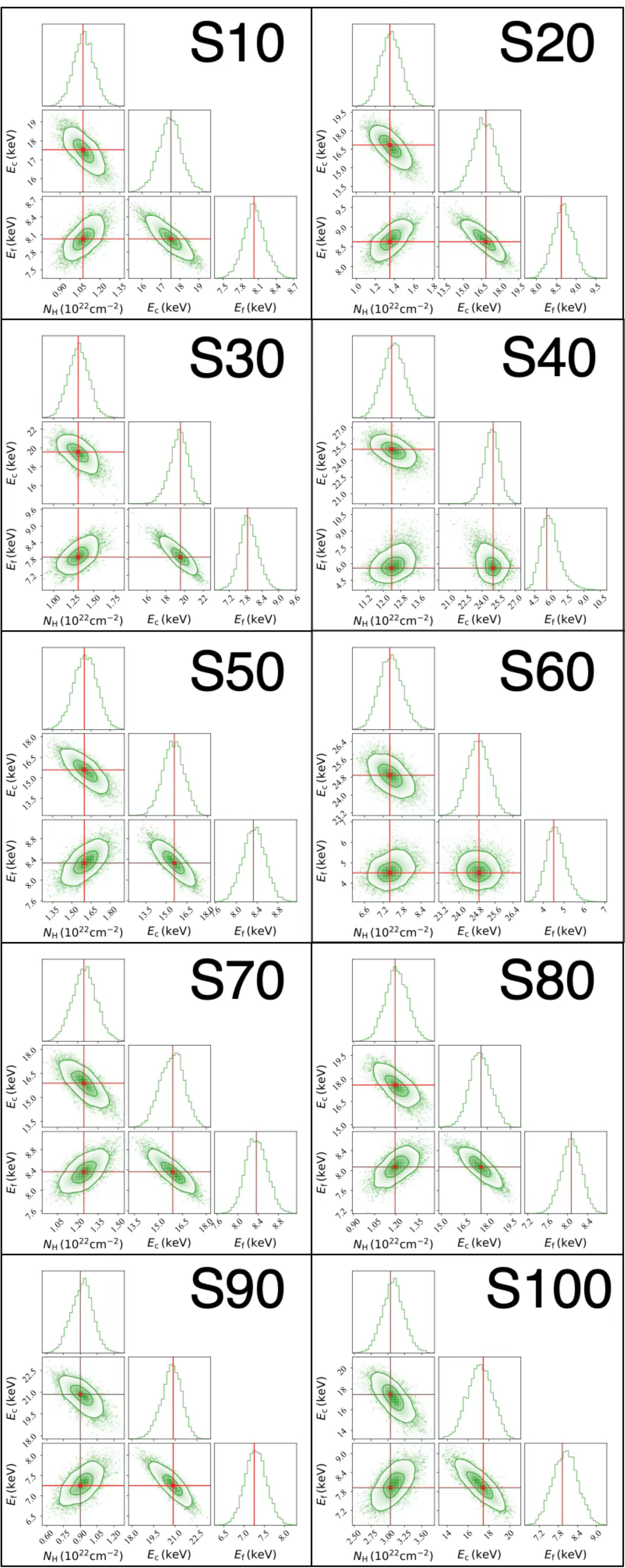}
    \caption{Corner plots of the fitting parameters $N_\mathrm{H}$, $E_\mathrm{c}$, and $E_\mathrm{f}$. Red and green lines indicate the best fit parameters and confidence regions ($1\sigma$ and $2\sigma$).}
    \label{fig:corner_plot}
\end{figure}

When fitting the continuum components of the Suzaku data described in Section~\ref{sec:spectral_analysis}, parameter degeneracies were examined using a Markov Chain Monte Carlo (MCMC) method.
We performed the MCMC analysis with the \textit{chain} command in \texttt{Xspec} (Fig.~\ref{fig:corner_plot}).
The Goodman--Weare algorithm \citep{Goodman2010} was used, with a burn-in of 10{,}000 steps and a total chain length of 30{,}000 steps.
Fig.~\ref{fig:corner_plot} shows the corner plots, which were produced by \textit{corner.py} \citep{Corner2016}, for $N_\mathrm{H}$, $E_\mathrm{c}$, and $E_\mathrm{f}$.

\bsp	
\label{lastpage}
\end{document}